\documentclass[12pt]{article}

\usepackage{graphics,psfrag}
\usepackage{amsthm,amssymb,epsfig,amsmath,euscript,array,cite}

\newcommand{\be}{\begin{equation}}
\newcommand{\ee}{\end{equation}}
\def\bea{\begin{eqnarray}}
\def\eea{\end{eqnarray}}
\newcommand{\eq}[1]{(\ref{#1})}
\def\nn{\nonumber}

\newcommand{\beq}{\begin{equation}}
\newcommand{\eeq}{\end{equation}}
\newcommand{\ben}{\begin{eqnarray}}
\newcommand{\een}{\end{eqnarray}}
\newcommand{\bes}{\begin{subequations}}
\newcommand{\ees}{\end{subequations}}
\newcommand{\blg}{\begin{align}}
\newcommand{\elg}{\end{align}}

\newcommand{\cN}{{\cal N}}

\newcommand{\startappendix}{
\setcounter{section}{0}
\renewcommand{\thesection}{\Alph{section}}}

\newcommand{\Appendix}[1]{
\refstepcounter{section}
\begin{flushleft}
{\large\bf Appendix \thesection: #1}
\end{flushleft}}

\def\one{\mbox{1 \kern-.59em {\rm l}}}

\def\a{\alpha}

  \def\D{\Delta}

\def\k{\kappa}
\def\l{\lambda} 
\def\m{\mu} 
\def\o{\omega}

\def\s{\sigma}  
\def\t{\tau}
\def\th{\theta}

 \def\cN{{\cal N}}

 \def\cZ{{\cal Z}}

\begin{document}

\hfill{WITS-CTP-050}

\vspace{20pt}

\begin{center}

{\Large \bf
On the non-BPS string solutions in Sasaki-Einstein gauge/gravity duality
}
\vspace{20pt}

{\bf
 Dimitrios Giataganas
}

{\em
National Institute for Theoretical Physics,\\
School of Physics and Centre for Theoretical Physics,\\
University of the Witwatersrand,\\
Wits, 2050,\\
South Africa
}

{\small \sffamily
dimitrios.giataganas@wits.ac.za
}

\vspace{30pt}
{\bf Abstract}
\end{center}
We present an extensive analysis on several string solutions in
$AdS_5\times Y^{p,q}$ and find some interesting properties of their energy-spin relations. Their energy depends always on the parameter $a(p,q)$ which characterizes these manifolds. The range of this parameter for the string solutions is constrained by the Sasaki-Einstein constraints that the solutions should satisfy. Hence some string solutions we find are not valid for the whole class of $Y^{p,q}$ manifolds. For some of our solutions, when the maximum allowed value of $a(p,q)$ corresponds to the  string approaching the poles of the squashed sphere in $Y^{p,q}$, their energy  at this limit approaches the BPS one. Thus certain non-BPS string solutions in the whole class of Sasaki-Einstein manifolds, can become BPS in particular manifolds. For the  solutions with this property we point out that this behavior is independent of the string motion in the other directions on the manifold. We expect that in the field theory the corresponding generic operators to these semi-classical strings, become BPS at certain quivers.

\setcounter{page}0
\newpage

\section{Introduction}

The last years there is a continuous effort to understand better the
duality between the type IIB superstring theory in $AdS_5\times S^5$ and
the $\cN=4$ supersymmetric Yang-Mills theory  \cite{maldacena1,witten1,gubser1}
and extend it to less
supersymmetric dualities. In this direction there are proposed several
gause/gravity dualities with less supersymmetries. One of these, are the
Sasaki-Einstein dualities, which contain a class of type IIB backgrounds
with  $\cN=1$ supersymmetry. They are of the form $AdS_5\times X^5$,
where $X^5$ is a Sasaki-Einstein manifold and are dual to $\cN=1$ superconformal
gauge theories called quivers.

A Sasaki-Einstein manifold is a positive curvature Einstein manifold whose
metric cone is Calabi-Yau. One can construct these manifolds by  probing
singular geometries using D-branes. More specifically using a stack of
$D3$-branes to probe singular Calabi-Yau we get 4-dimensional $\cN=1$ supersymmetric gauge
theories. In our case the multiple $D3$-branes  are placed at the singularity of
the metric cones over the Sasaki-Einstein manifolds. The geometry of the
singularity is responsible for the supersymmetry, the gauge group, the matter
 content and the superpotential interactions on the world volume of the D-branes.

There is a countably infinite class of Sasaki-Einstein metrics  $Y^{p,q}\cong S^2\times S^3$
\cite{gauntlettads,gauntletts2s3,gauntlettfinal} which is characterized by two relatively prime
integers $p,\,q$ with $q<p$. All these manifolds have a Reeb vector field which is a constant norm
Killing vector field and under the AdS/CFT correspondence is isomorphic to
the R-symmetry of the dual field theory. Depending on the global  properties
of the orbits of this Killing vector field, there is a way to classify the
Sasaki-Einstein manifolds to regular, quasi-regular and irregular manifolds.
Notice that these terms are related to the properties of the orbits of the
vector field and not to the regularity of the metric which is smooth.
In the first class, the roots of a cubic function $q(y)$ appear in the
metric of our spaces, are rational which implies that the parameter $a$,
which is related to $p$ and $q$ is also rational. For some of these values
of $a$ we get quasi-regular Sasaki-Einstein metrics and the corresponding
manifolds have the property that the orbits close but the associated $U(1)$
action is only locally free. It is however possible to have $p/q$ rational, even
in cases that the roots of $q(y)$ are irrational and then we speak for irregular
Sasaki-Einstein metrics where the orbits of the vector field do not close.

Since the isometries generated by the Reeb vector field on the Sasaki-Einstein
manifold are dual to the R-symmetry, the above classification should apply to
the properties of the R-charge. Hence, the regular and quasi-regular classes
are dual to theories with compact $R$-symmetry, where the R-charges of the fields
are rational. On the other hand, the irregular manifolds are dual to theories with non compact
$R$-symmetry and are invariant under the superconformal algebra but not the
superconformal group. This also means that the $R$-charges of the fields can
be irrational.

The dual field theories of these backgrounds can be constructed following
the fact that the Calabi-Yau cone of our manifolds is toric since the $U(1)^3$
group acts as a symmetry of $Y^{p,q}$. This is equivalent to the fact that there
are three conserved momenta in the $D3$-brane dynamics along the directions that
parametrize the three torus $\mathbb{T}^3$. The toric diagrams then can be computed
by introducing a moment map $\m:=C(Y^{p,q})\rightarrow \mathbb{R}^3$, where the image
is always a polyhedral cone, and the moment map exhibits $C(Y^{p,q})$ as a $\mathbb{T}^3$
fibration over this moment cone. It has been found that the cone is a four faceted
convex rational polyhedral cone, and the Delzant theorem can be applied
to obtain a gauged linear sigma model for $C(Y^{p,q})$. This is done in \cite{martelligeometry}
where the authors found that there is a $U(1)$ gauged linear sigma model with four
chiral superfields and charges $(p,p,-p+1,-p-q)$.
The most important to remember is that the dual field theory of the $Y^{p,q}$ background, has a product gauge group
$U(N)\times U(N)$, hence it has bifundamental matter superfields. Two of these fields
transform in the $(N,\bar{N})$ and the other two in the $(\bar{N},N)$ \cite{benvenuti04a}.

With all these details known, it is easy to understand that the basic properties
of the Sasaki-Einstein dualities, have been examined and understood well since some time ago.
The next step was to try to use these dualities as a tool to understand better the
theories with less supersymmetries and also to check the duality itself by
obtaining results in both sides. The results that were found firstly was for semiclassical
string solutions and their dual field theory operators for the special case of $AdS_5\times T^{1,1}$
examined in \cite{kim1,pons1,schvellinger1}, since the duality with the $T^{1,1}$
manifold was explained earlier in \cite{klebanov1}.
For the $Y^{p,q}$ manifolds a study of the BPS massless geodesics and their dual
long BPS operators has been done in \cite{benvenutiY}  and similar analysis applied
to the generalized spaces with cohomogeneity two, called $L^{p,q,r}$ \cite{cvetic1},
 in \cite{benvenuti4}.
The dual BPS giant gravitons have been studied in \cite{martelli2} and recently giant magnons
and spiky strings moving in a sector of $AdS_5\times T^{1,1}$ have been examined in \cite{benvenuti5}.
More recently a set of point-like and semi-classical extended string solutions examined in \cite{giataganasse1}.
Moreover, it is possible to marginally deform the theory \cite{benvenuti2}, and the case of
$\beta$-deformed Sasaki-Einstein dualities is examined in \cite{zaffaronibse}, where giant gravitons
are also analyzed. The giant magnons and single spike solutions were analyzed for a subsector of the deformed  $T^{1,1}$ in \cite{rashkovt11}.

It is very useful to try to understand the non supersymmetric sector of
states on $AdS_5\times Y^{p,q}$. In the original Maldacena's conjecture
 there is a good understanding already in supersymmetric and even for the
non-supersymmetric sector of states of
string theory on $AdS_5\times S^5$  \cite{bmn,gubser02,frolov02}.
In this case it is expected a close
connection between semiclassical string solutions and some integrable models,
since the corresponding $O(N)$ sigma models are known to be at least classically integrable.
For example, the folded rotating  string solutions with one or two non zero angular momenta
are related to 1-d sine-Gordon model \cite{devega,gubser02,frolovns,frolov02}
or the more generalized solutions with three non vanishing angular momenta are related
to the $1$-d Neumann model \cite{frolovintegr}. The technic consists of expanding
the energy of the rotating string solutions by assuming  the limits
$\l/J_i^2\ll 1$ and $1/J_i\ll 1$. Then one can compare the energy with the conformal
dimension in the super Yang-Mills side. An important remark here is that in the original gauge/gravity duality the superstring
$\a'$ corrections to the classical energy will be suppressed
by terms proportional to higher powers of $1/J_i$ and one reason for that is the global supersymmetry
of the underlying string theory even for the non-BPS solutions.
During this period, there was a lot of work on these topics and many papers were published finding and analyzing new close string solutions
and relating them to field theory operators, look for example \cite{tseytlin1} and references inside.

Therefore, as a first step would be useful to try to find some semiclassical extended string
solutions and their energy-spin relation in $AdS_5\times Y^{p,q}$. It seems that because of the
reduced global symmetry in these spaces, compared to the $S^5$, and due to the
constraints of the Sasaki-Einstein metrics, the process of finding these solutions is more difficult.
For example  one can find the solutions that solve the equations of motion and the Virasoro
constraints but they do not satisfy the Sasaki-Einstein constraints and hence have to be
discarded. In a previous paper \cite{giataganasse1} we started a discussion on the semiclassical strings on
Sasaki-Einstein spaces. We  presented an extended discussion on
the BPS solutions on the Sasaki-Einstein manifolds, and argue that do not exist any other than the one found in \cite{benvenutiY}.
We also tried to present the basic technic of finding extended string solution in the Sasaki-Einstein
spaces with cohomogeneity one and two, with the latter being the $L^{p,q,r}$ manifolds.
In this paper we extend our results and find  solutions for a very wide range of
string configurations. We check if the solutions we find are acceptable according to the Sasaki-Einstein constraints and
the periodic boundary conditions and many of them have to be discarded. We also find some new interesting
properties of the energy-spin relations and comment on them, with particular focus on their behavior
to approach the BPS energy in some limits.
 For our ansatze we activate a non-$U(1)$ angle and the three angles
along the $U(1)$ directions in the internal manifold. In $AdS$ part the string is localized
at a point and as usual the global time is expressed through the worldsheet time.

The paper is organized as follows. In section 2, we present the basic setup, which includes
all the equations that we have to solve and the conserved quantities we are going to use.
Moreover we explain the form of our ansatze. This section is supported by a more detailed
discussion  on the Sasaki-Einstein backgrounds and their constraints presented in the Appendix A.
In next section we find point-like and extended string solutions parameterized by the three $U(1)$ angles
and one angle that does not correspond to  a $U(1)$ direction. Moving to the next section we deactivate
the non-$U(1)$ angle and  try to generalize more the string motion on the three $U(1)$ directions.
This section is supported by the Appendix B.
For the solutions we find, we present their energy-spin relation and analyze their properties. In the last section we discuss
our results and make proposals for further analysis.

\section{String solutions in $Y^{p,q}$ background }
\subsection{Equations of motion and conserved quantities}

Here we present the basic setup we use to find the string
solutions in  $Y^{p,q}$ manifolds. We will write down the general
action, the corresponding equations of motion and the Virasoro constraints.
 The conserved charges are also presented, since are needed
to find the energy-spin relations.

First of all we give the basics of our background\footnote{For a more detailed discussion look at the appendix A.}
 starting with the
Sasaki-Einstein metrics $Y^{p,q}$ on $S^2\times S^3$  in the following local form \cite{gauntletts2s3}:
\ben
\label{se1}
  d s^2 &=& \frac{1-cy}{6}(d\theta^2+\sin^2\theta
      d\phi^2)+\frac{1}{w(y)q(y)}
      d y^2+\frac{q(y)}{9}(d \psi-\cos\theta  d \phi)^2 \nonumber\\
      & + &  {w(y)}\left[d \alpha +f(y) (d\psi-\cos\theta
      d \phi)\right]^2,
\een
where
\ben
w(y)  =  \frac{2(a-y^2)}{1-cy}~ ,\quad
q(y)  =  \frac{a-3y^2+2cy^3}{a-y^2}~ , \quad
f(y)  =  \frac{ac-2y+y^2c}{6(a-y^2)}~.
\een
For convenience we scale for the rest of the paper the constant $c$ to 1 by a diffeomorphism.

The ranges of the coordinates $(\theta,\phi,y,\psi)$ are $0\le\theta\le \pi$,
$0\le \phi \le 2\pi$, $y_1\le y\le y_2$ and $0\le \psi \le 2\pi$. The coordinate $\psi$ is the
azimuthal coordinate on the axially
squashed $S^2$ fibre and the round sphere $S^2$ parametrized by $(\theta,\phi)$.
The parameter $a$ in the metric depends on $p,\,q$ \eq{apq} and is restricted to the range
\be\label{ai}
0<a<1~.
\ee
By choosing the above range for $a$, the following conditions
of $y$ are satisfied: $y^2<a$, $w(y)>0$ and  $q(y) \geq 0$.

The range of $y$ is specified as follows. The equation $q(y)=0$ is cubic
and has three real roots, one negative and two positive. Naming the negative
root $y_{q-}$ and the smallest positive root $y_{q+}$ we must choose the range of the coordinate $y$ to be
\be\label{yqy}
y_{q-}\leq y\leq y_{q+}~,
\ee
with the boundaries corresponding to the south and north poles of the axially  squashed $S^2$ fibre $(\psi,\,y)$.
Also, it is necessary to have $p/q$ rational in order to have  a complete manifold. More explicitly the following equality holds
\ben\label{xi}
y_{q+}-y_{q-}=\frac{3 q }{2 p}\equiv \xi~,
\een
where $\xi$ is defined for later use.
Notice also that for $y=y_{q\pm}$ we get two three-submanifolds and the cones over them  are calibrated with respect to four-form $\frac{1}{2} J\wedge J$, where $J$ is the K\"{a}hler form. Hence they are supersymmetric.

We can now write the general action for our system. From the $AdS$ part of the space we use only the global time which expressed through the world-sheet time as $t=\k \t$, and the string is localized at the point $\rho=0$. The Polyakov action\footnote{For an interesting discussion on the given name of the action look at \cite{polyakov}} in the conformal gauge is given by
\ben\nonumber
S&=&-\frac{\sqrt\l}{4 \pi}\int d\t d\s\big[- (-\dot{t}^2+t'{}^2)+\frac{1-y}{6}(-\dot{\theta}^2+\theta'{}^2)+\frac{1}{w q}(-\dot{y}^2+y'{}^2) \\\nonumber
&&+(\frac{1-y}{6}s_\th^2+\frac{q}{9}c_\th^2+w f^2 c_\th^2)(-\dot{\phi}^2+\phi'{}^2)+(\frac{q}{9}+w f^2)(-\dot{\psi}^2+\psi'{}^2) +w(-\dot{\a}^2+\a'{}^2)\\\nonumber &&
-2 c_\th(\frac{q}{9}+w f^2)(-\dot{\psi}\dot{\phi}+\psi'{}\phi'{})+
2 w f(-\dot{\a}\dot{\psi}+ \a'{}\psi'{})-2 w f c_\th(-\dot{\a}\dot{\phi}+\a'{}\phi'{})\big]~.
\een
Where for convenience we do not write explicitly the dependence of $y$ in the functions $f,\,w$ and $q$.
The classical equations of motion for $\theta$ and $y$ take the form
\ben\nonumber
&&\frac{1-y}{6}(-\dot{\phi}^2+\phi'{}^2)s_{2\theta} + ( \frac{q}{9}+ w f^2)\big(s_{2\theta}(\dot{\phi}^2-\phi'{}^2)+2 s_\theta(-\dot{\psi}\dot{\phi}+\psi'{}\phi'{})\big)\\\label{eom1}
&&\hspace{2.7cm}+2 w f s_\theta(-\dot{\a}\dot{\phi}+\a'{}\phi'{})+\frac{1-y}{3}(\ddot{\th}-\th'')
+\frac{1}{3}(-\dot{\th}\dot{y}+y'\th')=0~,\\
\nonumber
&&\frac{s_\theta^2}{6}(\dot{\phi}^2-\phi'{}^2)+(\frac{Q}{9}+ A_1 )\Big(c_\theta^2(-\dot{\phi}^2+\phi'{}^2)-\dot{\psi}^2+\psi'{}^2-2 c_\theta(-\dot{\psi}\dot{\phi}+\psi'{}\phi'{})\Big)+W(-\dot{\a}^2+\a'{}^2)\\
&&+2 A_3\Big(-\dot{\a}\dot{\psi}+ \a'{}\psi'{}+c_\theta(\dot{\a}\dot{\phi}-a'{}\phi'{})\Big)+\frac{1}{6}(\dot{\th}^2-\th'{}^2)+\frac{2}{w q}(\ddot{y}-y'')+2 A_2(\dot{y}^2-y'^2) =0~,\label{eom2}
\een
while for the $U(1)$ angles $\a,\,\phi$ and $\psi$ read
\ben\nonumber
&&w(\a''-\ddot{\a})+w f (\psi''-\ddot{\psi})- w f c_\th(\phi''-\ddot{\phi})+W(a' y'-\dot{a}\dot{y})\\
&&\hspace{2.5cm}+A_3(\psi' y'-\dot{\psi}\dot{y})- A_3 c_\th(\phi'y'-\dot{\phi}\dot{y})+ w f s_\th(\phi'\th'-\dot{\phi}\dot{\th})=0~, \label{eom3}\\\nn
&&-w f c_\th(-\ddot{\a}+\a'')+(\frac{1-y}{6}s_\th^2+\frac{q}{9}c_\th^2+w f^2 c_\th^2)(-\ddot{\phi}+\phi'')+\\\nn
&&-c_\th(\frac{q}{9}+ w f^2)(-\ddot{\psi}+\psi'')+s_\th(\frac{q}{9}+ w f^2)(-\dot{\psi}\dot{\th}+\psi'\theta')+ w f s_\theta(-\dot{\a}\dot{\th}+a'\th')\\\nn
&&+(\frac{1-y}{6}-(\frac{q}{9}+ w f^2))s_{2\th}(-\dot{\th}\dot{\phi}+\theta' \phi') +(-\frac{s_\th^2}{6}+(\frac{Q}{9}+ A_1) c_\th^2)(-\dot{y}\dot{\phi}+ y' \phi')\\
&&\hspace{3.7cm}-A_3(-\dot{a}\dot{y}+a'y')c_\th-c_\th(\frac{Q}{9}+ A_1)(-\dot{\psi}\dot{y}+\psi' y')=0~, \label{eom4}
\een
\ben\nn
&&w f(-\ddot{\a}+\a'')-c_\th(\frac{q}{9}+ w f^2)(-\ddot{\phi}+\phi'')+(\frac{q}{9}+ w f^2)(-\ddot{\psi}+\psi'')+(\frac{Q}{9}+A_1)(-\dot{\psi}\dot{y}+\psi'y')\\
&&\hspace{1cm}+A_3(-\dot{a}\dot{y}+\a'y')+(\frac{q}{9}+ w f^2)s_\th(-\dot{\th}\dot{\phi}+\th'\phi')+((\frac{Q}{9}+ A_1)c_\th(-\dot{y}\dot{\phi}+y'\phi')=0~,
\label{eom5}
\een
where the following conventions have been used
\be
A_1:=\partial_y(w f^2),\quad A_2:=\partial_y((w q)^{-1}),\quad A_3:=\partial_y(w f),\quad Q:=\partial_y q,\quad  W:=\partial_y{w}.
\ee
Since we use the Polyakov action, there are two further equations that need to be satisfied. The Virasoro equations read
\ben\nonumber
&&\frac{1-y}{6}s_\theta^2\dot{\phi}\phi'{}+(\frac{q}{9}+ w f^2 )\big(c_\theta^2\dot{\phi}\phi'{}+\dot{\psi}\psi'{}-c_\theta(\dot{\phi}\psi'{}+\dot{\psi}\phi'{})\big)
+w \dot{\a}\a'{}\\
&&\hspace{2cm}+w f \big(\dot{\a}\psi'{}+\dot{\psi}\a'{}-(\dot{\a}\phi'{}+\dot{\phi}\a'{})c_\theta\big)+\frac{1-y}{6}\dot{\th}\th'+\frac{1}{w q}\dot{y}y'=0,\label{vc1}\\\nonumber
&&-\k^2+\frac{1-y}{6}s_\theta^2(\dot{\phi}^2+ \phi'{}^2)+ w (\dot{\a}^2+ \a'{}^2)+2 w f \big(\dot{\a}\dot{\psi}+\a'{}\psi'{}-(\dot{\a}\dot{\phi}+\a'{}\phi'{})c_\theta\big)
\\\nonumber
&&\qquad\qquad\qquad\quad\quad+
(\frac{q}{9}+ w f^2 )\big(c_\theta^2(\dot{\phi}^2+ \phi'{}^2)+\dot{\psi}^2+ \psi'{}^2
-2 c_\theta(\dot{\phi}\dot{\psi}+\phi'{}\psi'{})\big)
\\
&&\qquad\qquad\qquad\qquad\qquad\qquad\qquad\,\,\,\,\,\,+\frac{1-y}{6}(\dot{\theta}^2+\theta'{}^2)+\frac{1}{w q} (\dot{y}^2+y'{}^2)=0 ~.
\label{vc2}
\een
The manifold $Y^{p,q}$ has three $U(1)$ symmetries and hence admits three conserved charges which are the angular momenta corresponding to strings rotating along the $\a,\,\phi$ and $\psi$ directions which presented below:
\ben\label{1ja}
J_a&=&\frac{\sqrt{\l}}{2 \pi}\int_{0}^{2 \pi}d \sigma\,(w \dot{a}-  w f c_\theta \dot{\phi} +w f \dot{\psi})~,\\\label{1jphi}
J_\phi&=&\frac{\sqrt{\l}}{2 \pi}\int_{0}^{2 \pi}d \sigma\,\big( -w f c_\theta \dot{a}+(\frac{1-y}{6}s_\theta^2  + \frac{q}{9}c_\theta^2 +w f^2 c_\theta^2)\dot{\phi}-(\frac{q}{9} +w f^2)c_\theta \dot{\psi}\big),\\\label{1jpsi}
J_{\psi}&=&\frac{\sqrt{\l}}{2 \pi}\int_{0}^{2 \pi}d \sigma\,\big( w f \dot{a}-(\frac{q}{9} +w f^2)c_\theta\dot{\phi}+(\frac{q}{9} +w f^2)\dot{\psi}\big)~.
\een
There exists one more conserved quantity, the classical energy, which is generated by the translational invariance along $t$
and can be written as
\ben\label{1e}
E&=&\frac{\sqrt{\l}}{2 \pi}\int_{0}^{2 \pi}d \sigma\, \k ~.
\een
Finally, an other expression we are going to use is the $R$-charge which is related to the momenta $J_a$ and $J_\psi$ as
\be\label{Rc}
Q_R=2 J_\psi-\frac{1}{3}J_\a~.
\ee

\section{String solutions in $Y^{p,q}$ background}

In this section we present some spinning string solutions  in the $Y^{p,q}$ manifold.
In the most general case the stings can be spatially extended in the direction 
$\theta$, spin and extend along the three $U(1)$ directions of $Y^{p,q}$ and rest in the other directions.
More specifically it is allowed
to string to move on a circle of the round sphere $S^2$ parametrized by the coordinate $\phi$. Moreover, on the squashed sphere the string can move on its azimuthal coordinate $\psi$, and sit at a constant value $y_0$ between the north and south poles. Furthermore, the string can move on the principle $S^1$ bundle over $B$ \eq{metricse2} parametrized by $\a$.  In the $AdS$ space the string is localized at the point $\rho=0$  and the global time is expressed through the world-sheet time as $t=\k \t$.

The described motion of the string can be parametrized by the following ansatz
\ben\nn
&&\a=\o_1\t+m_1 \s,\qquad\phi=\o_2\t+m_2 \s,\qquad\psi=\o_3\t+m_3 \s~,\\\label{ansatz2}
&&t=\k\t,\qquad\theta=n\s\qquad\mbox{and}\qquad y=y_0,
\een
where $y_0$ is constant to be determined and the numbers $n,\,m_i$ must  be integers due to the periodicity condition in the global coordinates of the manifold on $\sigma$.
Our purpose is to use this ansatz to find new semiclassical solutions, and their energy in terms of the momenta.

\subsection{Extended string in $\theta$ with spin along the $U(1)$ directions}

Let us  start by allowing the string to spin along the $\a$ direction, which can be  done by using the ansatz
\be
\th=n\s,\qquad\a=\o_1\t~.
\ee
The only equation that is not trivially satisfied is \eq{eom2} which solved easily to specify $y$ as
\be\label{lysh1}
y=1\pm 2 \sqrt{3}\sqrt{\frac{ 1-a}{ n^2+12 \o_1^2}}\,\o_1~.
\ee
It is obvious that only the solution with the minus sign could be acceptable since $y<1$.
The solutions contains three free parameters, so it is complicated to check analytically whether our solution satisfy the constrain \eq{yqy}. It is possible however to reduce temporarily the parameters to two, by defining $n= k \o_1$, where $k$ can be any number and choose without any loss of generality $\o_1>0$. Then the solution \eq{lysh1} becomes
\be\label{lysh1k}
y=1-2 \sqrt{3}\sqrt{\frac{ 1-a}{12+k^2}}
\ee
and can be seen in the Figure 1 \footnote{In the whole paper we use the convention that the solutions of $q(y)$ $y_{q-},\,y_{q+}$, are plotted with red and green colors respectively. The relevant solution $y$ to the equations of motion and the Virasoro constraint is plotted as blue.}, that there are infinite values of $k$ for which the constrain \eq{yqy} is satisfied.
Actually, one can even find the analytic expression for the boundary values $k$ in terms of $a$, by equating the expressions $y_{q+}$ and $y$ from \eq{lysh1k}. The boundary of the acceptable values for k is presented in the second plot of Figure 1.
\begin{figure}\label{flysh1}
\includegraphics[width=70mm]{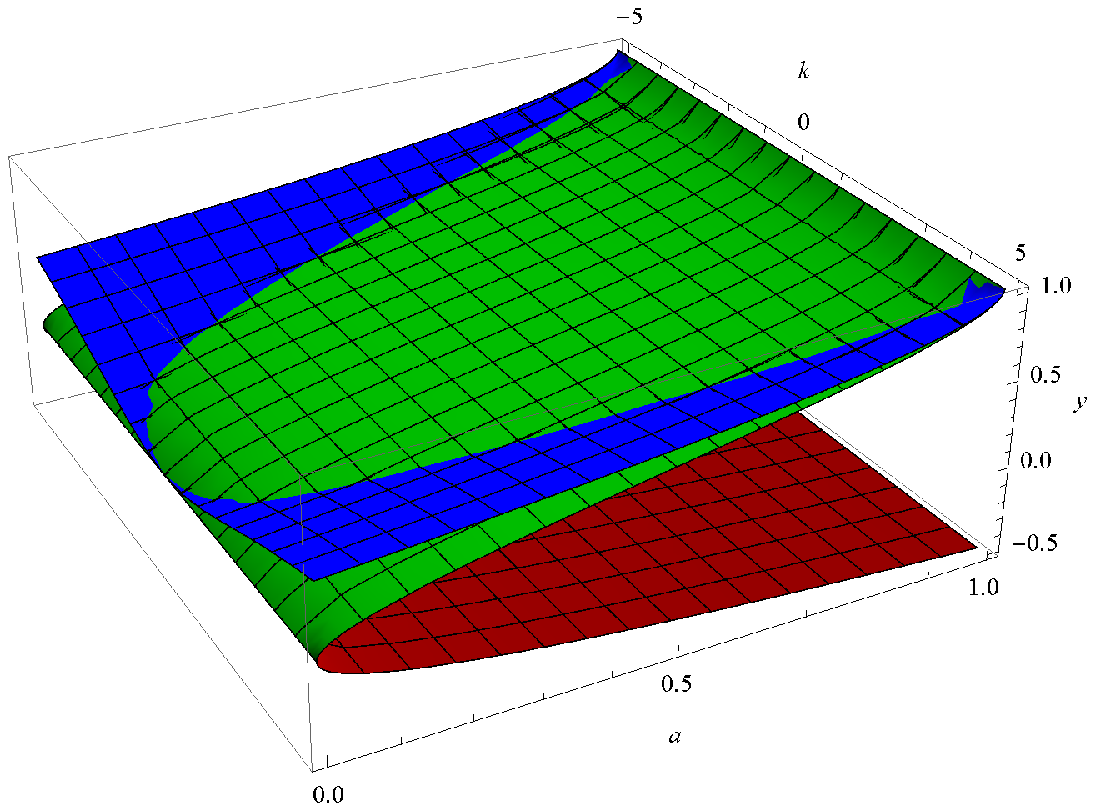}
\includegraphics[width=70mm]{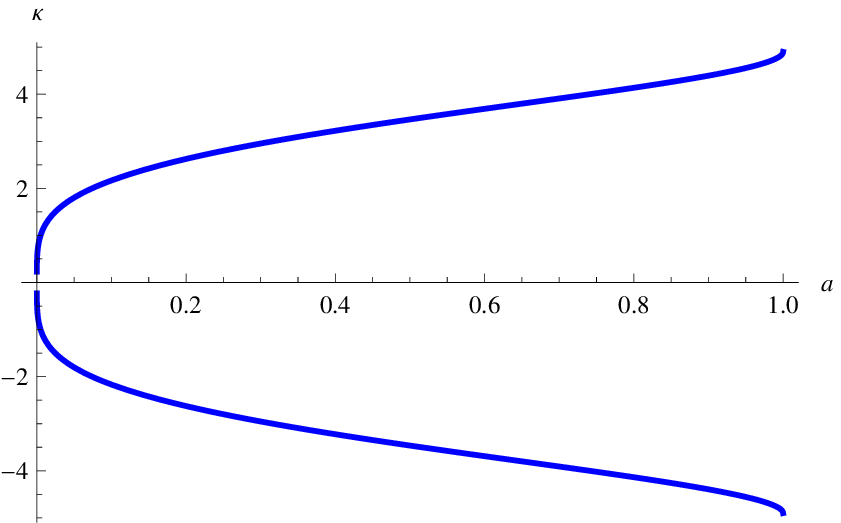}
\caption{{\small In the first plot of the that  we see the solution $y$ \eq{lysh1k} compared with $y_{q\pm}$ plotted with running parameters  $a$ and $k$. There is an acceptable region for a wide range of values of $k$. For the color conventions look the footnote 3. In the second plot is presented the solution  $k(a)$ of the equation $y=y_{q+}$. The minimum and maximum acceptable values of $k$ are approximately $\pm 4.899$ for $a\rightarrow 1$.}}
\end{figure}
Hence, the Sasaki-Einstein constraints are setting limits on the relation between the parameters $n,\,\o_1$. Since we know that
there exist acceptable solutions we restore our initial ansatz and continue our analysis.

To get the energy of our solution, we should look at the second Virasoro constraint  which  gives
\be\label{lysh1kk}
\k^2=\frac{1-y}{6}n^2 +w \o_1^2=4 y \o_1^2~,
\ee
where for the second equality we used the equation \eq{eom2}.
The corresponding momenta are given by the following simple expressions
\be\label{lysh1jj}
J_\a=\sqrt{\l}w \o_1, \qquad J_\phi=0, \qquad J_\psi=\sqrt{\l}w f \o_1~.
\ee
There are many ways to express the energy in terms of the momenta. A convenient one is to use the R-charge $Q_R$ which produce a relatively compact result when inserted in Virasoro constraint \eq{lysh1kk}
\be
\k^2=\frac{9}{4} \frac{Q_R^2}{y} \,
\ee
where we also used \eq{Rc}.
Here $y$ is a constant but depends on $\o_1$ and $n$. To find a relation that depends only on the conserved quantities we can divide the conserved quantities and by using the equations \eq{lysh1jj} we write $y$ as
\be
y= \frac{1-\sqrt{1-a+36 a J_{\psi/\a}^2}}{1+6 J_{\psi/\a}},
\ee
where $J_{\psi/\a}:= J_\psi/J_\a$. So the result can be written as
\be\label{energeia1}
E=\frac{3 }{2}|Q_R|\sqrt{\frac{J_\a+6 J_\psi}{J_\a- \sqrt{(1-a)J_\a^2+36 a J_\psi^2}}}
\ee
This relation depends both on $J_a$ and $J_\psi$. In the limit $a\rightarrow 1$ where $J_a\rightarrow 4 \o_1$ and $J_\psi\rightarrow 0$, the energy approaches $3/2Q_R$ which is the energy of the point-like BPS solutions.

As a next step let us consider a string that spins in $\phi$ direction, and extend along $\th$. A parametrization of this motion is
\be
\th=n\s,\qquad\phi=\o_2\t~,
\ee
 and the resulting non-trivial equations of motion are \eq{eom1} and \eq{eom2}. The first equation is solved only for $\th=0,\,\pi/2,\,\pi$,
 supposing as usual $a\neq 1$, which has a result that the string is localized at $\th$. This case was examined in \cite{giataganasse1} where found that has no real solution.
Finally, in this class of simple spinning strings, we consider a motion of a string parametrized by the following ansatz
\be\label{ansatz3}
\th=n\s,\qquad\psi=\o_3\t~.
\ee
Only the equation of motion for $y$ is not trivially satisfied which gives the solution
\be\label{lysh3}
y=1\pm \frac{\sqrt{3}}{3} \sqrt{\frac{1-a}{n^2-\o_3^2}} \, \o_3 ~,
\ee
with $n^2>\o_3^2$. We need to check if the Sasaki-Einstein constraints are satisfied and for this reason we reduce the parameters in the above expression by setting temporarily
$n= k \o_3$  and considering $\o_3>0$ which simplifies the solution as
\be\label{klysh3}
y=1\pm\frac{\sqrt{1-a}}{\sqrt{3} \sqrt{\left(k^2-1\right)}}~.
\ee
By equating that to $y_{q\pm}$ we see that there is a narrow interval for $k$ where the constrain \eq{yqy} is satisfied. The boundaries of this interval as functions $k(a)$ can be easily found, but the relations are quite lengthy and it is better to present the results in the Figure 2, since the exact function is not important at present. It is however interesting to  remark that the approximate maximum range of the intervals for acceptable $k$ appear for $a\rightarrow 1$ and are
\be\label{kkl}
k\in \pm(1,\,1.414)~.
\ee
\begin{figure}\label{flysh3}
\includegraphics[width=70mm]{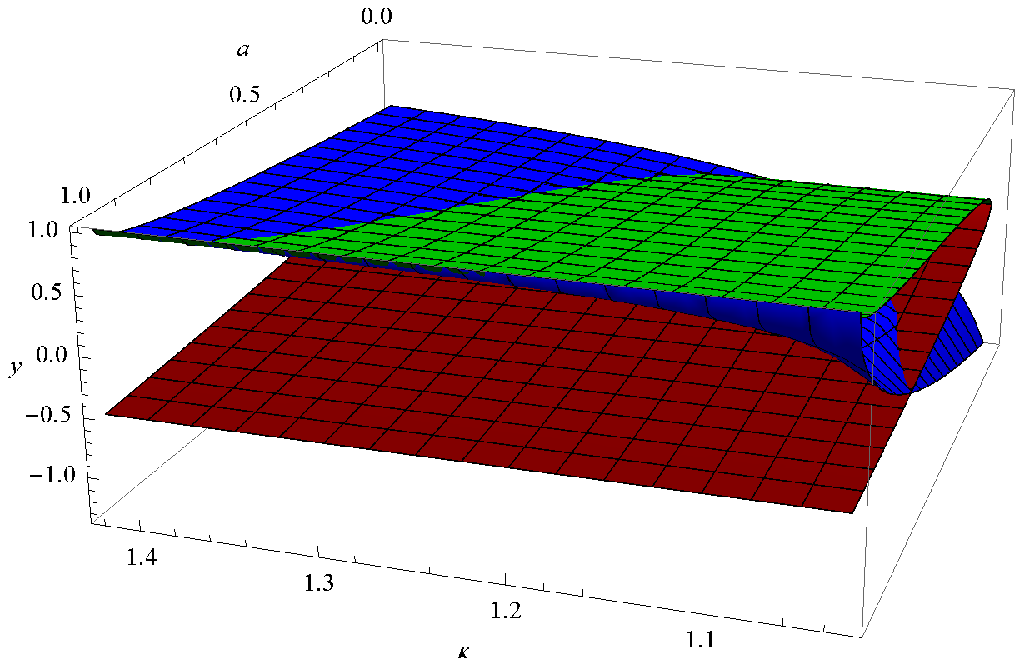}
\includegraphics[width=70mm]{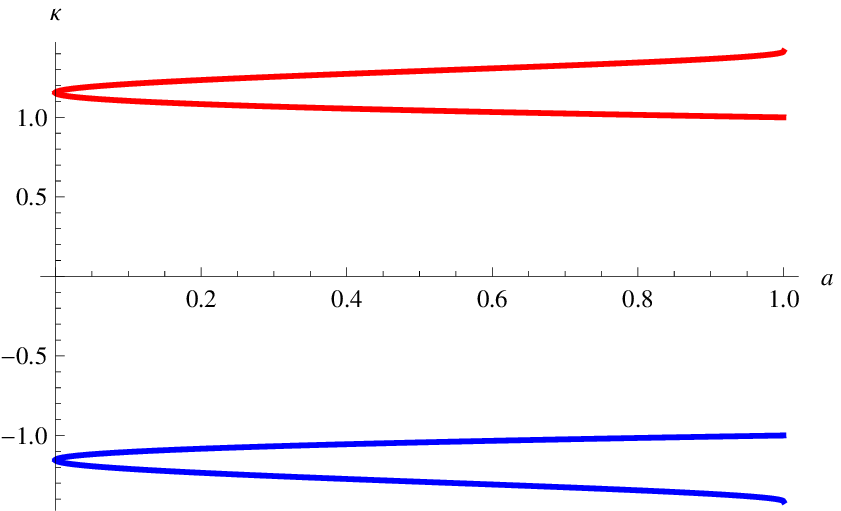}
\caption{{\small In the first plot, notice that the solution $y$ \eq{klysh3} is between  $y_{q\pm}$ for some values of  $a$ and $k$. In the second plot is presented the solution  $k(a)$ of the equation $y=y_{q\pm}$.  There are two solutions, symmetric to $k=0$ axis. Above the $k$-axis the minimum and maximum acceptable values of $k$ are approximately $1$ and $1.414$ for $a\rightarrow 1$.}}
\end{figure}
Hence the non zero conjugate momenta using the acceptable solution \eq{lysh3},  are
\ben\label{momenta3}
J_\a=\sqrt{\l} \frac{\sqrt{1-a}  \left(4 \o_3^2-3 n^2\right)\o_3}{3 \sqrt{3} \sqrt{  \left( n^2-\o_3^2\right) \o_3^2}},
\qquad J_\psi=\sqrt{\l}  \frac{\sqrt{(1-a)}  \left(2 \o_3^2- n^2\right)\o_3}{6 \sqrt{3} \sqrt{  \left( n^2-\o_3^2\right) \o_3^2}}~. \qquad
\een
The Virasoro constraint \eq{vc2}, using the equations of motion give
\ben\label{lysh3kk}
\k^2=\frac{\sqrt{1-a} \o_3^4}{3 \sqrt{3} \sqrt{  \left(n^2-\o_3^2\right)\o_3^2}}=\frac{1}{3}(1-y)\o_3^2=\frac{27}{4(1-y)} Q_R^2 ~,
\een
where to write the last equality we used the fact  that
\be\label{qrlysh3}
Q_R=\sqrt{\l} \frac{2(1-y)}{9}\o_3~.
\ee
It is obvious that we need to relate $y$ with the momenta in order to find the energy spin relation. Dividing $J_\psi$ with $J_\a$ and using the relations \eq{momenta3}, we can solve for $y$ to get
\be\label{ylysh3}
y=1-\frac{\sqrt{(1-a) (1+4 J_{\psi/\a} (3 J_{\psi/\a}-2))}}{\sqrt{3} (1-2 J_{\psi/\a})}~,
\ee
where $y$ is chosen from the quadratic equation that arose, such that it satisfies the constraint \eq{yqy}. To do that we should notice that $J_{\psi/\a}$ written in terms of $k$  is always negative in the interval \eq{kkl}. 
Now using  the equations \eq{lysh3kk} and \eq{qrlysh3} we can find
\ben\label{energeia3}
E=\frac{3}{2} |Q_R| \sqrt{\frac{3 \sqrt{3}(J_\a-2 J_\psi)}{\sqrt{(1-a)(J_\a-6 J_\psi)(J_\a-2 J_\psi)}}}~.
\een
The energy depends on $J_a$ and $J_\psi$ and comparing it with the  relation \eq{energeia1} we notice that there is a common factor of $3/2 Q_R$ in both relations.

\subsection{Motion of the string in two $U(1)$ directions}

In this section we examine several string solutions that are allowed to move only along the $U(1)$ directions. Hence in the initial ansatz \eq{ansatz2} we
change $\th=\th_0$ where $\th_0$ is a constant angle. Then we are classifying almost all the acceptable solutions that produced from the ansatz \eq{ansatz2}. However there are some few cases that the computational work is demanding and the final result lengthy and complicated and we choose not to present them here.

The choice of the angle $\th_0$ is a tricky issue and will be chosen equal to $\pi/4$. This choice simplifies the equations since  $\sin\th$ and $\cos\th$ are equal and at the same time keep relatively all the dynamics of the Sasaki-Einstein manifolds, ie. does not make zeroth  any terms in the equations of motion and the Virasoro constraints. An other choice would be to set $\th=\pi/2$, but this has a result of several vanishing terms in the equations we need to solve, and hence we do not capture the dynamics of the motion in our spaces in a full generality. Therefore in the rest of the paper we are going to use
\be
\th=\pi/4
\ee
and only when is stated differently the angle $\th$ will take other value.

Moreover, the advantage of considering for the string motion the linear ansatz \eq{ansatz2} along the three $U(1)$ directions and require it to rest at the remaining ones, is that the three equations of motion \eq{eom3}, \eq{eom4}, \eq{eom5}, corresponding to the three $U(1)$ angles, are trivially satisfied.

We start by turning on the angles $\a,\,\psi$ where also the BPS point-like string is allowed to move \cite{benvenutiY}.
We are allowing to the string to spin and extend along the $\a$ direction and extend spatially in $\psi$ as\footnote{We are looking for such solutions that have frequencies and winding numbers on the ansatz considered not equal to zero. This is a convention we will use in the whole paper. For example for this ansatz we present the solutions for $\o_1$ and $m_1,\,m_3$ not equal to zero.}
\be
\a = \o_1 \t+ m_1 \sigma,\qquad \psi = m_3 \sigma
\ee
for which the non trivial equations \eq{eom2} and \eq{vc1} solved by
\be\label{lysh5}
m_1 = \frac{m_3}{6},\, \qquad\, a = \frac{2 m_3^2}{m_3^2 + 9 \o_1^2},\, \qquad \, y = a~.
\ee
Hence the winding number $m_3$ should be multiple of six. Furthermore the inequality \eq{yqy} constrain $m_3$ (Figure 3) as
\be\label{lysh5in}
|m_3|<\sqrt{3}\, \o_1 \Rightarrow a<\frac{1}{2}~,
\ee
\begin{figure}\label{flysh5}
\includegraphics[width=70mm]{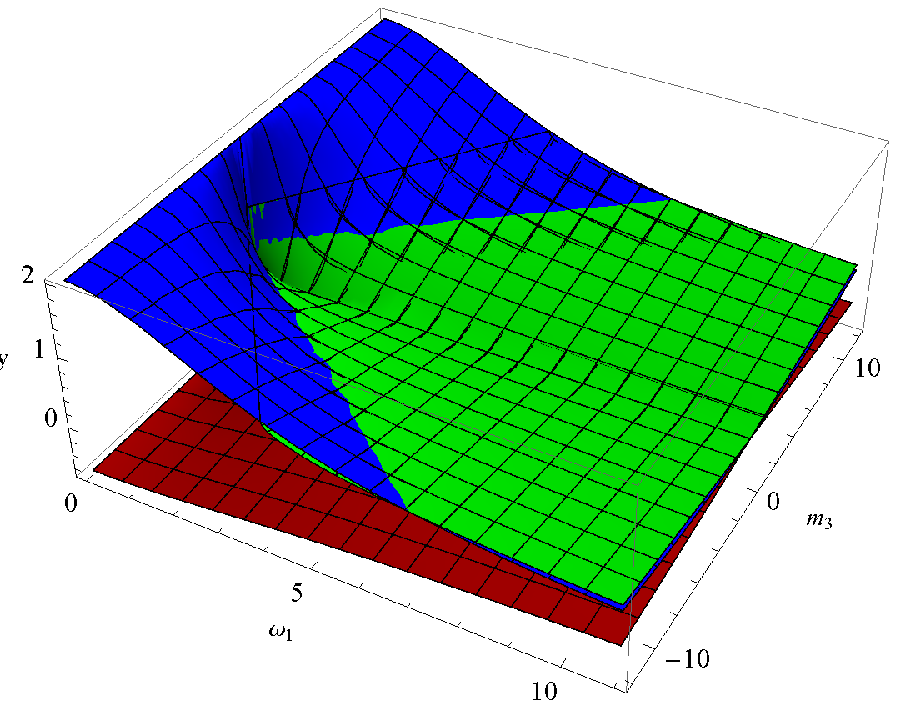}
\includegraphics[width=70mm]{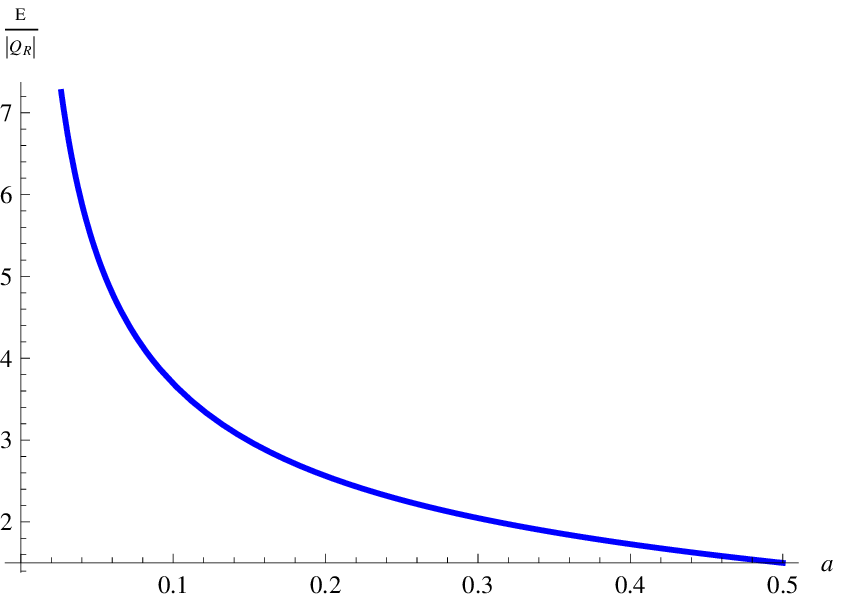}
\caption{{\small In the first plot, we see the solution $y$ \eq{lysh5} compared with $y_{q\pm}$ plotted with running parameters  $m_3$ and $\o_1$. The acceptable region is for $a<1/2$ specified in \eq{lysh5in}. Notice that $m_3$ is allowed to take only the values that are multiples of six, however we plot the whole surface to see the behavior of the function. In the second plot we see the factor of proportionality in the energy relation \eq{lysh5energy} versus $a$. This function is monotonically decreasing with respect to $a$ with minimum value $3/2$ for $a\rightarrow 0.5$ and as we approach $a\rightarrow 0$ it is rapidly increasing.}}
\end{figure}
where we consider $\o_1$ non-negative.   In this region the second Virasoro constraint \eq{vc2} for the corresponding solution becomes
\be\label{twok212}
\k^2=\frac{\left(15 \o_1^2-m_3^2\right)m_3^2 }{3 \left(m_3^2+9 \o_1^2\right)}~,
\ee
where the R.H.S is always positive in the interval \eq{lysh5in}. To write the energy in terms of the momenta we need to find their expressions which  are given by
\ben\label{twomom2122}
J_\a=\sqrt{\l} \frac{4 m_3^2 \o_1}{m_3^2 + 9 \o_1^2},\qquad J_\phi=\sqrt{\l} \frac{\sqrt{2} m_3^2 \o_1}{3 \left(m_3^2+9 \o_1^2\right)},\qquad
J_\psi=-\sqrt{\l} \frac{2 m_3^2 \o_1}{3 \left(m_3^2+9 \o_1^2\right)},
\een
and are related each other as
\be\label{ppp12}
J_\a=-6 J_\psi=6 \sqrt{2} J_\phi=-\frac{3}{2} Q_R~,
\ee
where we have used the definition of the R-charge.

For our solution it is obvious that the requirement for rationality of $y_{q+}-y_{q-}$ is satisfied since the parameter $a$ given by the relation \eq{lysh5} can take continuous values. It is better however, to express $\o_1$ in terms of $a,\, m_3$ like
\be\label{lysh5o1}
\o_1=\pm\frac{\sqrt{2 m_3^2-a m_3^2}}{3 \sqrt{a}}=\frac{\sqrt{\left(-4 \xi^2 \sqrt{9-3 \xi^2}+3 \left(9+\sqrt{9-3 \xi^2}\right)\right) m_3^2}}
{3 \sqrt{9-3 \sqrt{9-3 \xi^2}+4 k^2 \sqrt{9-3 \xi^2}}}~.
\ee
The second expression for $\o_1$ comes by  using the equation \eq{axi} for $a$. The expression \eq{lysh5o1} have the advantage that we can choose values of $a$ or $\xi$ such that the relation $y_{q+}-y_{q-}$  is rational and for these acceptable values we would specify the parameters of our ansatz. Notice that $\xi$ is constrained by the inequality \eq{lysh5in} as
\be
\xi<\frac{\sqrt{3}}{2}~.
\ee
which is also sufficient to keep $\o_1$ real.
This interval is a subset of the interval where $\xi$ lives.
Here we will carry on working the equation $\o_1(a){}$\footnote{Obviously this equation will replace the second equation of \eq{lysh5}.} of \eq{lysh5o1} since it is also more convenient to express the energy directly in terms of $a$. Notice also that we could work with the parameter $\xi$ instead of $a$ but the equations become
lengthy.

For our solution the second Virasoro constraint and the R-charge $Q_R$ take the simple form
\ben\label{virn12}
\k^2=\frac{1}{9} (5-4 a) m_3^2,\qquad Q_R=-\sqrt{\l} \frac{4}{9} \sqrt{(2-a)a m_3^2}~,
\een
with $J_\psi,\,J_\phi$ and $J_\a$ can be found from \eq{ppp12}.
Then the energy in terms of the momenta read
\ben\label{lysh5energy}
E=\frac{3}{4} \sqrt{\frac{5-4 a}{(2-a) a}}|Q_R|~.
\een
The energy is proportional to $Q_R$ and the factor of proportionality is a monotonically decreasing function with respect to $a$. The minimum value of this factor is for $a\rightarrow 1/2$, where it is approaching $3/2$ which is exactly the value that the BPS solutions have \cite{benvenutiY,giataganasse1}. Hence for values of $a$ close to $1/2$ the energy of our solutions is approaching the energy of the BPS states. It is interesting to point out that the solutions we are currently examining, have energy just above the BPS solutions because the range of $a$ is limited to $(0,1/2)$. The parameter $a$ is constrained in order the solution \eq{lysh5} to satisfy the Sasaki-Einstein constraint \eq{yqy}. This an interesting remark, the Sasaki-Einstein constraints
ensure that the minimum energy of this solution can not go below the energy of the BPS solutions.

Let us choose some specific manifolds as  examples starting with the manifold $Y^{2,1}$, which have $a\simeq 0.387 \Leftrightarrow \xi=3/4$, where we have used the equation \eq{apq}.
The energy of our solution in this manifold is
\ben
E=\frac{2}{3} \sqrt{\frac{2 \left(18536+1523 \sqrt{13}\right)}{6873}} |Q_R|\simeq 1.763 |Q_R|
\een
while for $Y^{3,1}$, which have $a\Leftrightarrow 0.181 \Leftrightarrow \xi=1/2$ and
\ben
E=\frac{3}{16} \sqrt{\frac{3}{29} \left(1143+151 \sqrt{33}\right)}|Q_R| \simeq 2.704 |Q_R| ~.
\een
The second manifold differ for one unit in $q$ compared to the first but its parameter $a$ is almost the half of $Y^{2,1}$.
However even for the $Y^{3,1}$, $a$ is far away enough from zero, and hence the energies of the two manifolds do not have big gap. By considering manifolds with $a$ close to zero we would see steepest changes in their energies.


An other sensible ansatz is to allow the string to spin according to the following relations
\be
\a = m_1 \sigma,\qquad \psi =\o_3 \t+ m_3 \sigma ~.
\ee
Here the non trivial equations are \eq{eom2} and \eq{vc1} and are solved for
\ben\label{lysh6}
\o_3&=&\sqrt{\frac{m_3 (m_3+6 m_1)^2}{m_3 + 4 m_1}}~,\\\label{lysh6y}
y&=&1\pm\frac{\sqrt{(1-a) (2 m_1+m_3) (6 m_1+m_3)}}{\sqrt{3} (2 m_1+m_3)}~,
\een
where $m_1$ and $m_3$ can take integer values while $\o_3$ is chosen to satisfy $\o_3>0$. We can choose $\o_3$ and $m_i$  positive without loss of generality. Doing that, the $y$ solution with the minus sign is smaller than one and can be acceptable. To check if there is a region that the solution is acceptable we can consider temporarily  $m_1= k m_3$, where $k$ is a positive integer. This results the reduction of free parameters in \eq{lysh6y}  and allow as to compare easier with $y_{q\pm}$ since we get
\be\label{yklysh6}
y=1-\frac{\sqrt{(1-a) (1+2 k) (1+6 k)}}{\sqrt{3} (1+2 k)}~,
\ee
and it can be easily seen that the solution is acceptable for a wide choice of $k$, with the results are also presented in Figure 4. Since we now know that there exist acceptable solutions we can continue the analysis with our initial general ansatz.
\begin{figure}\label{flysh6}
\centerline{\includegraphics[width=70mm]{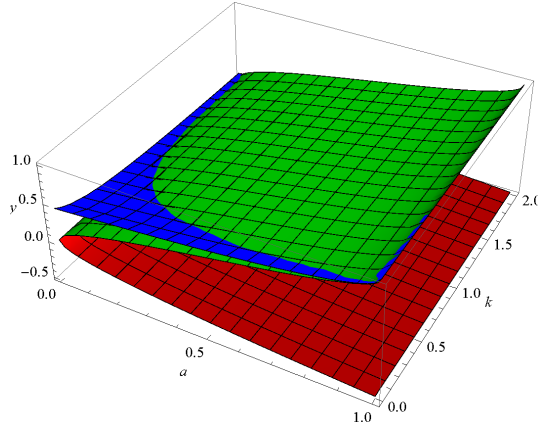}}
\caption{{\small The figure present the comparison of the solutions \eq{yklysh6} and $y_{q\pm}$. It is obvious that the constraint \eq{yqy} is satisfied for a wide range for values of $k$. Hence there exist acceptable $y$ solutions.}}
\end{figure}

In order to find the energy we must look at \eq{vc2} which provide the following lengthy expression
\ben\label{energylysh6}
\k^2=\frac{4 m_1^2 \left(72 m_1^2+54 m_1   m_3 -2\sqrt{3}\left(6 m_1+  m_3 \right)\sqrt{(1-a) (2 m_1+m_3) (6 m_1+m_3)}+9 m_3^2\right)}{9 (2 m_1+m_3) (4 m_1+m_3)}.
\een
Furthermore the conjugate momenta are given by
\ben
J_\a=\sqrt{\l} w f \o_3=-\sqrt{\l}\frac{2 m_3^{3/2} \sqrt{(1-a) (6 m_1+ m_3)}}{3 \sqrt{3} \sqrt{(2 m_1+ m_3) (4 m_1+ m_3)}}~,\\
J_\psi=-\frac{m_1}{m_3} J_\a,\qquad J_\phi=\frac{m_1}{\sqrt{2} m_3} J_\a~,\qquad Q_R=-\frac{6 m_1 +m_3}{3 m_3}J_\a~.
\een
We obtained at least  one simple expression relating the fraction of the winding numbers and the fraction of the momenta. Although it is possible to express the energy in terms of the momenta, it appears to be difficult to find a relatively compact way to write this relation. The energy-spin relation here is  lengthy and transcendental  due to the complicated way that the solutions depend on the winding numbers. So we decide not to write down the energy for this solution in terms of the momenta.

As final step we take the most general ansatz of the type considered:
\be
\a = \o_1 \t+ m_1 \sigma,\qquad \psi =\o_3 \t+ m_3 \sigma~.
\ee
The equations \eq{eom2} and \eq{vc1} give several solutions. However the most of them do not satisfy the rationality condition \eq{xi} with integer winding numbers.  To show how the solutions look we pick the simplest one:
\be
\o_1 = \frac{\o_3}{2},\qquad  m_3 = \o_3,\qquad  m_1 = \frac{m_3}{2},\qquad   y = \frac{1}{9}(1 + 8 a)~.
\ee
By checking the constraint \eq{yqy} we see that the function $y$ is greater than  $y_{q+}$  and becomes equal to this only for $a = 5/32$ or $a=1$. There are several other reasons not to accept these solutions. One can see that for this value of $a$ the fraction $p/q$ becomes equal to $1/2\sqrt{1/2 \left(3-\sqrt{5}\right)}$, thus $p,\,q$ can not be integer numbers. Similar is the situation for the rest of the solutions we have checked, and it seems that a general acceptable solution for this ansatz do not exist.

Before finishing the examination of the string motion in these two angles we mention that the following ansatze
\ben
&&\a = \o_1 \t,\qquad \psi =m_3 \sigma,\\
&&\a = m_1 \sigma,\qquad \psi =\o_3 \t~,
\een
do not have any acceptable solutions and this is the reason that we did not present them in the beginning of our analysis.
Moreover, there are several other string solutions with motion allowed to these two directions examined in \cite{giataganasse1}.

We can  now move to another class of string solutions by activating two other angles $\a$ and $\phi$.
We begin with the simpler point-like ansatz
\be\label{ansatz8}
\a = \o_1 \t,\qquad \phi = \o_2 \t,
\ee
which gives the non trivial equations \eq{eom1}, \eq{eom2}. These are solved for
\ben\label{lysh8}
\o_1=\frac{\o_2}{3 \sqrt{2}},\qquad y_{\pm}=1\pm\frac{\sqrt{1-a}}{\sqrt{2} }~,
\een
where we consider $\o_2>0$ and the only solution that could be acceptable is the one with the minus sign since it is smaller than $1$.
This solution satisfies the constraint \eq{yqy} for $a>1/2$  and plotted in the Figure 5.
\begin{figure}\label{flysh8}
\includegraphics[width=70mm]{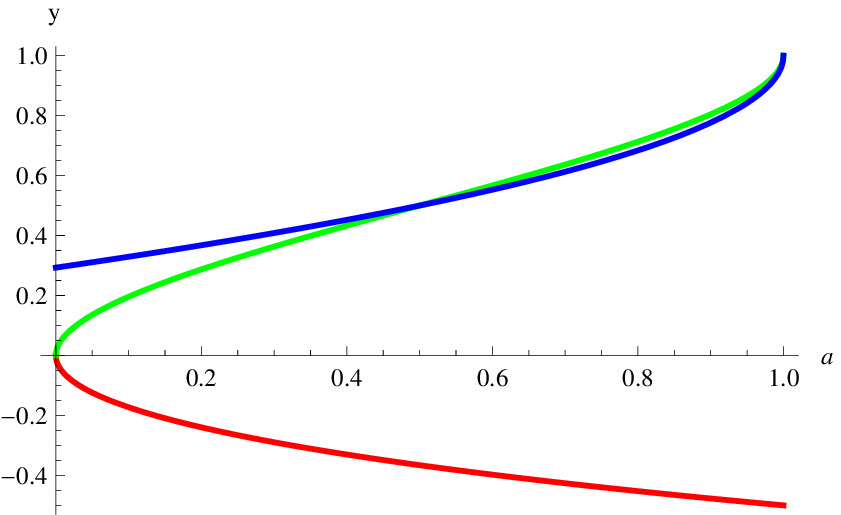}
\includegraphics[width=70mm]{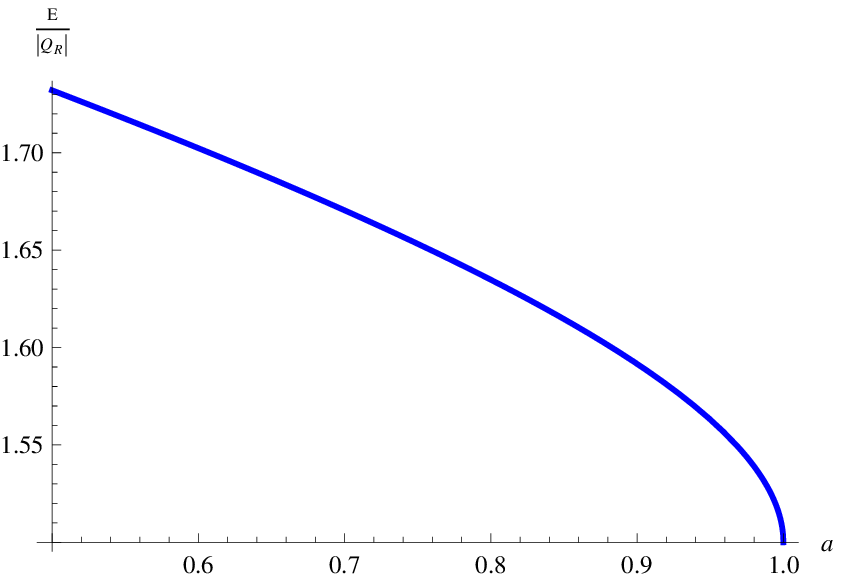}
\caption{{\small In the first plot we notice that the solution $y$ \eq{lysh8} is between $y_{q+},\, y_{q-}$ for $a>1/2$. In the second plot is presented the energy divided by $|Q_R|$, and we see that the curve is always above $3/2$ as expected, and approaches this value for $a\rightarrow 1$.}}
\end{figure}

The second Virasoro constraint \eq{vc2}  now gives
\be\label{enlysh8}
\k^2=\frac{1}{18} \o_2^2 \left(4 -\sqrt{2} \sqrt{1-a}\right)~,
\ee
and the conserved charges are
\ben
J_\a=\frac{\sqrt{\l}}{6} \left(4 \sqrt{2} -5 \sqrt{1-a}\right)\o_2,\quad
J_\phi=\sqrt{\l}\frac{\sqrt{1-a} }{6 \sqrt{2}}\,\o_2,\quad
J_\psi=-\frac{\sqrt{\l}}{12} \sqrt{1-a}\,\o_2~,
\een
where $J_\a,\,J_\phi$ are  positive, while $J_\psi$ is negative and related each other and to the R-charge by
\be
J_\phi=-\sqrt{2}J_\psi=\frac{\sqrt{1-a}}{8-5 \sqrt{2} \sqrt{1-a}}J_\a=\frac{3 \sqrt{1-a}}{2(\sqrt{2}\sqrt{1-a}-4)}Q_R~.
\ee
Hence the energy expressed in terms of the momenta and more specifically the R-charge takes the form
\be\label{energeialysh8}
E=\frac{3}{2}\sqrt{\frac{ 2 \left(4+\sqrt{2}\sqrt{1- a}\right)}{7+a}}|Q_R|~,
\ee
where the factor of proportionality is always greater that $3/2$, as expected since these solutions are not supersymmetric\footnote{The situation here is different with the massless geodesics in $S^5$, where are  all protected. In the Sasaki-Einstein spaces only a particular set of massless geodesics is BPS.} and approaches this value for $a\rightarrow 1$, as can also be seen in Figure 5. This is a common behavior with other cases we examined, where for $a\rightarrow 1$ the energy of the solutions approaches the BPS energy.

For completeness we mention that for the static string configuration $\a = m_1\sigma$ and  $\phi = m_2\sigma$,
the non trivial equations of motion are not solved for integer winding numbers since the relation $m_1=\frac{m_2}{3 \sqrt{2}}$ is necessary.

A generalization of the previous ansatz can be done by extending the string spatially in $\a$ direction as
\be\label{ansatz9}
\a = \o_1 \t+ m_1\sigma,\qquad \phi = \o_2 \t~,
\ee
where the string can spin in both directions parametrized by $\a$ and $\phi$.
In this case all the equations are non trivial and the solutions we find are
\ben
m_1&=&\pm\frac{1}{2} \sqrt{-12 \o_1^2+3 \sqrt{2} \o_1 \o_2-\frac{\o_2^2}{3}}~,\\
a&=&\frac{\o_2^2 \left(18 \sqrt{2} \o_1+\o_2\right)}{-1728 \sqrt{2} \o_1^3+18 \sqrt{2} \o_1 \o_2^2+\o_2^3},\qquad
y=\frac{\o_2}{12 \sqrt{2} \o_1+\o_2}~.
\een
The frequencies are constrained by the requirement that $m_1$ is real. In the acceptable region of frequencies, we get $a<0$ always and hence the solutions are not acceptable. This ansatz does not give any new solution.

We can modify the previous ansatz by eliminating the $\sigma$ dependence of the angle $\a$ and inserting it in the angle $\phi$, which gives
\be\label{ana3122}
\a = \o_1 \t,\qquad \phi = \o_2 \t+ m_2\sigma~.
\ee
Now, we end up with all equations to be non trivial. Their solutions have complex $m_2$ and we do not examine them further.

Before we analyze more generalized  string motion, we  briefly mention that the ansatze
\ben
&&\a = \o_1 \t,\qquad \phi = m_2\sigma~,
\\&&\a = m_1 \sigma,\qquad \phi = \o_2 \t~,
\een
do not give any new solutions to the equations of motion and the Virasoro constraints.

Finally, we consider the general ansatz which allows the string to spin in both directions and parametrized by
\ben
\a = \o_1 \t+m_1\sigma,\qquad \phi = m_2\sigma + \o_2 \t~.
\een
Now  all the equations are non trivial and more complicated. However they have some simple solutions which are checked and we found that do not satisfy the Sasaki-Einstein constraints. One example of this kind is
\ben
m_2=s_1\o_2,\quad m_1=s_1\o_1,\quad
\o_1=\frac{1}{12} \sqrt{2}\left(s_2\sqrt{7}- 1 \right)\o_2,\quad
y=\frac{5+\left(11-s_2 4 \sqrt{7}\right) a}{4 \left(4-s_2 \sqrt{7}\right)}
\een
where $s_{1,\,2}=\pm 1$. The $y$ solution is always greater that $y_{q+}$ and approach this boundary only for $a\rightarrow 1$. Hence it is not acceptable. There are
some other solutions that are very lengthy and complicated and due to the large amount of computational work we were not able to check if they satisfy the manifold constraints. From some special cases we were able to look, it seems that they do not satisfy them. It would be interesting to do however  an extensive check.

Therefore, it is worthy to try to simplify our system for that case by modifying the value of the constant angle $\theta$. We notice that when we set $\theta=\pi/2$
the equations we have to solve are simplified significantly compared to the ones above. It is not so difficult to see that  the equations of motions and the first Virasoro constraint are solved by
\be\label{lysh9}
m_2=\o_2,\qquad a=1-\frac{576 m_1^2 \o_1^2}{\left(\o_2^2-24 m_1 \o_1\right)^2},\qquad
y=\frac{\o_2^2}{\o_2^2-24 m_1 \o_1}~,
\ee
where we have many free parameters and is difficult to compare $y$ with $y_{q\pm}$. We can temporarily set $\o_2=1$ and then we can see that the constraint \eq{yqy} is satisfied (Figure 6). Actually there is a wide range for values of $\o_i,\,m_i$ for which the solutions satisfy the Sasaki-Einstein constrains, and hence it makes sense to continue our analysis for generic $\o_2$.
\begin{figure}\label{flysh9}
\centerline{\includegraphics[width=70mm]{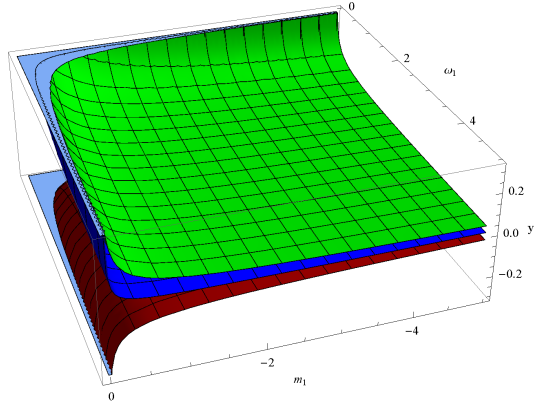}}
\caption{{\small We plot the solution $y$ \eq{lysh9} for $\o_2=1$ and $\o_1,\,m_1$ running parameters. It can be seen that the inequality $y_{q-}<y<y_{q+}$ is satisfied. }}
\end{figure}
It is not difficult to calculate the conserved momenta which are equal to
\be\label{lysh9mom1}
J_\a=\sqrt{\l}\frac{4\o_1 \o_2^2}{\o_2^2-24 m_1 \o_1}=-\frac{\o_2}{m_1}J_\phi=-3 Q_R~, \qquad J_\psi=0~.
\ee
From the second Virasoro constraint using the solution \eq{lysh9} we get
\be\label{lysh9ene1}
\k^2=\frac{4 ( m_1-\o_1)^2 \o_2^2}{\o_2^2-24  m_1 \o_1}~.
\ee
We can express the energy in terms of the momenta and the winding numbers $m_1$ using the following relations
\be
\o_2=-m_1\frac{J_\a}{J_{\phi}}~,\qquad \o_1=\frac{m_1 J_\a^2}{4(m_1 J_\a+6 J_\phi^2)}~,\qquad \k^2=-\frac{3(m_1-\o_1)^2}{\o_1}Q_R
\ee
which are derived with the use of the equations \eq{lysh9mom1} and \eq{lysh9ene1}. Hence the energy can be written in terms of the momenta as
\be
E=\frac{|J_\a^2-24 J_\phi^2-4  J_\a m_1|}{2}\sqrt{\frac{m_1}{J_\a(6 J_\phi^2+J_\a m_1)}}
\ee
The dispersion relation  is complicated, and the winding $m_1$ appears in such a way in the relation, because of the way the parameter  $a$ and the coordinate $y$ depend on it.

We have found several different solutions allowing the string to move on directions parametrized by the angles $\a,\,\psi$ and $\a,\,\phi$. To finish this section we will consider string motion in the remaining  two $U(1)$ directions parametrized by the angles
$\phi$, and $\psi$.

We begin with the simple point-like string
\be\label{ansatz10}
\phi = \o_2 \t,\qquad \psi = \o_3 \t,
\ee
for which the equations of motion do not give any new real solutions. The situation is similar for the following ansatze
\ben\label{anaf3ab122}
&&\phi = m_2 \sigma,\quad \psi = \o_3 t,\\
&&\phi = \o_2 t,\quad \psi = m_3 \sigma~,
\een
which do not give any new solutions as well as the non interesting static  ansatz $\phi = m_2 \sigma,\quad \psi = m_3 \sigma$.

However, by modifying the ansatz to
\be\label{anstatz11}
\phi = \o_2 \t +m_2\sigma,\qquad \psi = \o_3 \t,
\ee
we end up with all the three non trivial equations which are solved for
\ben\nn
\o_2&=&\frac{(9 - \sqrt{17}) \o_3}{8 \sqrt{2}},\qquad m_2=\frac{s_2}{8} \sqrt{11 \sqrt{17}-3} |\o_3|~,  \\\label{lysh11}
y&=&1-\frac{s_3 s_4}{6}\sqrt{3(3+ \sqrt{17})} \sqrt{1-a } ~,
\een
where $s_{2,3}=\pm 1$ and $s_4=\frac{|\o_3| }{ \o_3}=\pm 1$. Without loss of generality we choose $\o_3>0$, which suggests that $s_3=+1$ in order to have $y<1$. The sign of the variable $s_2$ does not matter so far and to avoid to carry it with us, we set it equal to $+1$. The constraint \eq{yqy}
is satisfied for $a>0.271$ as can also be seen in the Figure 7.
Notice also that $\o_3$ has to be multiple of $(1/8 \sqrt{11 \sqrt{17}-3})^{-1}$ is order to ensure that $m_2$ is integer.

To proceed to the conserved quantities we substitute our solutions to the second Virasoro constraint which gives
\be\label{klysh11}
\k^2=\frac{1}{24} \sqrt{\frac{5}{3}+\sqrt{17}} \sqrt{1-a} ~\o_3^2~.
\ee
The non zero conserved charges for the solution \eq{lysh11} are
\ben
J_\a&=&-\sqrt{\l}\frac{\sqrt{\left(51 \sqrt{17}-107\right) (1-a)}~ \o_3}{48 \sqrt{3}}\\
J_\psi&=&\sqrt{\l}\frac{\left(30+18 \sqrt{17}-\sqrt{3663+1929 \sqrt{17}} \sqrt{1-a}\right) (1-a) \o_3}{144 \left(4 \sqrt{3 \left(3+\sqrt{17}\right)} \sqrt{1-a}-\left(15+\sqrt{17}\right)(1- a)\right)}~.
\een
Both charges depend only on $\o_3$, and conveniently the $Q_R$ charge can be expressed in a short relation as
\be
Q_R=\frac{\sqrt{\l}}{72} \sqrt{\frac{109}{3}+9 \sqrt{17}} \sqrt{1-a} ~\o_3~
\ee
and as a result $Q_R$ takes discrete values too.
By taking the square root of the expression \eq{klysh11} and substituting $\o_3$ from the above equation we find the energy-spin relation which takes the simple form
\be
E=\frac{  3^{3/4} \sqrt{2} ~6}{\left(1397+339 \sqrt{17}\right)^{1/4}}\frac{Q_R}{(1-a)^{1/4}}.
\ee
As we expect the multiplication factor is greater than $3/2$ for any value of $a$ and is plotted in the Figure 7. However for this solution the energy does not approach the BPS one, even when $y\rightarrow y_{q\pm}$.
\begin{figure}\label{flysh11}
\includegraphics[width=70mm]{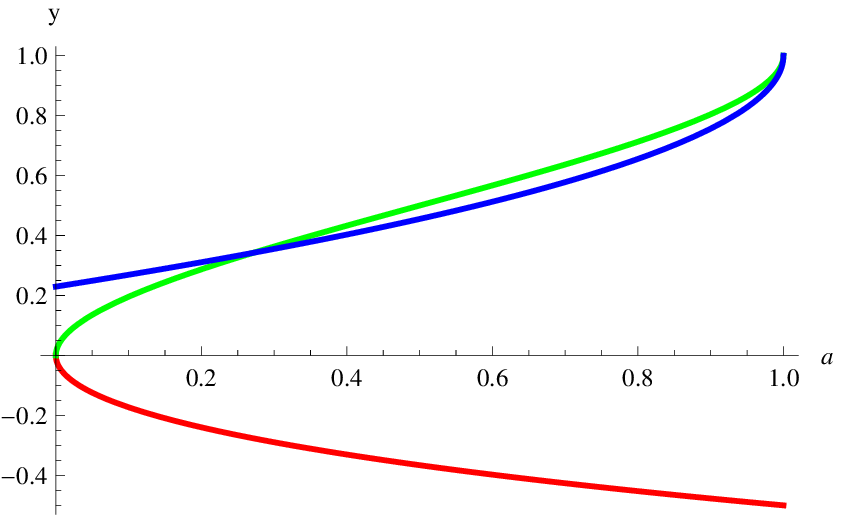}
\includegraphics[width=70mm]{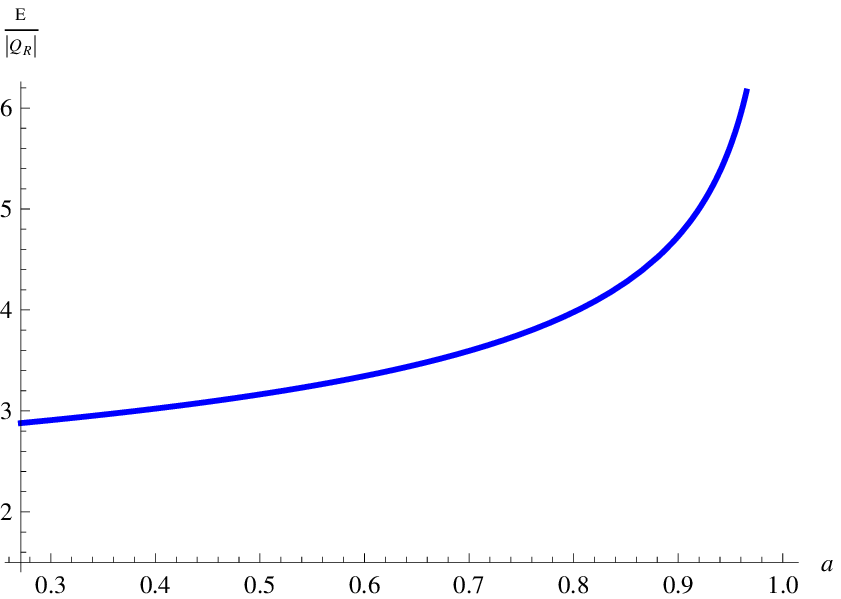}
\caption{{\small In the first plot we see the solution $y(a)$ \eq{lysh11} and its comparison with the $y_{q\pm}$. For $a>0.271$ inequality \eq{yqy} is satisfied and this is the interval we are interested in. In the second graph we see the behavior of the function $E/Q_R$ depending on $a$ which is a monotonically increasing function with lower bound
$2.879$ and when $a\rightarrow 1$ it increases rapidly. }}
\end{figure}


An interesting situation arise by considering an extended
string along the directions $\phi$ and $\psi$  which spins only along $\phi$ as
\be\label{ansatz11}
\phi = \o_2 \t +m_2\sigma,\qquad \psi = m_3 \s.
\ee
In case that the solutions to the relevant equations relate $\o_2$ to $m_i$, then the whole configuration is constrained a lot. We must ensure that the winding numbers are both integers and this could require
fixed form of $\o_2$. However to have both integer winding numbers, related to one frequency, is not always possible and this happens for the solutions we are about to examine.

Inserting the ansatz \eq{ansatz11} in the equation of motion and the first Virasoro constraint we get the following solutions
\ben
m_3=\sqrt{\frac{3+11 \sqrt{17}}{32}} \, \o_2,\quad m_2=\frac{1}{2} \sqrt{\sqrt{17}-3}\, \o_2,\quad
y=1-\frac{1}{6} \sqrt{3(3+\sqrt{17})} \sqrt{1-a}
\een
Where we already required $\o_2>0$ without loss of generality. The solution for $y$ is equal to the solution \eq{lysh11} and hence satisfies the inequality \eq{yqy}.
However, in order to ensure that $m_2$, is an integer we require  $\o_2= n\,2(\sqrt{17}-3)^{-1/2}$, where $n$ is an integer.
Then  the winding $m_3$ clearly is not an integer. So these solutions are not  interesting.

To avoid the previous situation we can consider an ansatz like
\be\label{ansatz12}
\phi = \o_2 \t, \qquad \psi = \o_3 \t +m_3\sigma,
\ee
which include two frequencies and one winding number.
Inserting it to  the relevant equations of motion and Virasoro constraints gives no new real solutions.

Finally, we are inserting $\sigma$ and $\t$ dependence in both angles we are considering the motion of the string as
\be\label{ansatz14}
\phi = \o_2 \t+m_2\sigma,\qquad \psi = \o_3 \t +m_3\sigma.
\ee
The solutions for this case are complicated. Due to the large amount of computational work we were not able to check if all of them satisfy the manifold constraints. As in the case of the solutions of strings moving along $a$ and $\phi$, we are simplifying the initial ansatz by considering the constant angle $\th=\pi/2$. Thus the equations of motion are simplified significantly and their solutions are
\be
y=\frac{1}{3} \left(3-\sqrt{3} \sqrt{1-a}\right),\quad \o_2=\pm \sqrt{2} \sqrt{-\o_3^2+m_3^2}~.
\ee
Unfortunately the $y$ solution is always greater than $y_{q+}$  and approach the pole of the squashed sphere at the limit $a\rightarrow 1$. Hence there is no string solution for $\theta=\pi/2$ with the ansatz \eq{ansatz14}.

So far we have examined extensively the motion of the string along two $U(1)$ isometries. In the next section we will try to find possible string solutions for strings moving along three $U(1)$ directions.

\subsection{Motion of the string in  three $U(1)$ directions}

In this part we are activating for the motion of the string all the three angles that parametrize the $U(1)$ directions. We start again from the simplest ansatz and try gradually to examine more complicated configurations.
We start by considering the point-like string as
\be\label{ansatz15}
\a=\o_1 \t,\qquad\phi = \o_2 \t,\qquad \psi = \o_3 \t,
\ee
and setting again the $\theta$ angle equal to $\pi/2$ in order to simplify the already complicated equations. Then the nontrivial equations are \eq{eom1} and \eq{eom2} which have solutions
\be\label{lysh15}
\o_1=-\frac{\o_2^2+2 \o_3^2}{12 \o_3},\qquad y=1-\frac{\sqrt{1-a} \o_2}{\sqrt{\o_2^2-4 \o_3^2}}~.
\ee
We have to impose the inequality
\be\label{inlysh15}
\o_2^2>4 \o_3^2,
\ee
 in order for the solution $y$ to take real values and we can also consider $\o_2>0$. By using the same trick  we set temporarily $\o_2=k \o_3$ in order to reduce the variables in function $y$ and we plot it in terms of $k$ and $a$ so to compare with the solutions of $q(y)$, $y_{q\pm}$ (Figure 8). We can see that there is a very wide range of $k$ values where our solution is acceptable. Hence we are allowed to continue to calculate the conjugate momenta which read
\ben
J_\a&=&-\frac{(\o_2^2+2 \o_3^2)( \o_2^2-4 \o_3^2)-\o_2^3 \sqrt{(1-a) \left(\o_2^2-4 \o_3^2\right)}}{3 \o_3(\o_2^2 -4 \o_3^2)},\\
J_\phi&=&\frac{\sqrt{1-a}~ \o_2^2}{6  \sqrt{\o_2^2-4 \o_3^2}}~,
\een
while $J_\psi=0$. It seems there is so simple way to insert the conserved charges in the energy which is the square root of the RHS of the second Virasoro constraint \eq{vc2}, and takes the following form after substituting the solutions of the equations of motion
\be\label{klysh15}
\k^2= \frac{\o_2^4+4 \o_2^2 \o_3^2+4 \o_3^4-\o_2^3 \sqrt{(1-a) \left(\o_2^2-4 \o_3^2\right)}}{36 \o_3^2}~.
\ee
The mathematical relations after expressing $\o_2,\,\o_3$ in terms of the momenta and inserting them in the energy \eq{klysh15} become very lengthy and we choose not to present them here. We point out that they can be found analytically but are  lengthy and transcendental.

\begin{figure}\label{flysh15}
\centerline{\includegraphics[width=70mm]{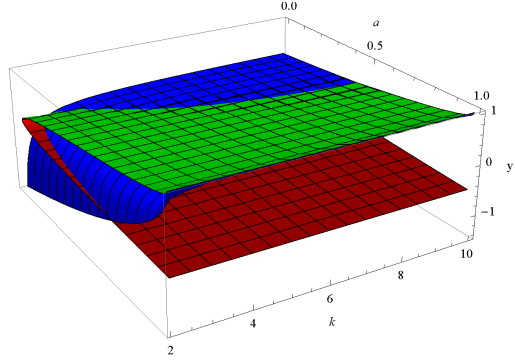}}
\caption{{\small The solution \eq{lysh15} is plotted for $\o_2=k \o_3$ and compared with $y_{q\pm}$. Clearly there is an infinite interval of $k$ that the solutions satisfy the inequality \eq{yqy}.}}
\end{figure}

As a brief remark we mention that for the static ansatz
\be\label{ansatz16}
a=m_1 \sigma,\qquad \phi = m_2 \s,\qquad \psi = m_3 \s,
\ee
for $\th=\pi/4$  the corresponding solutions exclude integer winding numbers and are not acceptable.

It is not difficult to see that many  spinning strings  extended only along one $U(1)$ direction and spin along all the three $U(1)$ directions do not solve the equations of motion and the Virasoro constraints. For example by considering
\be\label{ansatz17}
a=\o_1 \t,\qquad \phi = \o_2 \t,\qquad \psi = \o_3 \t+ m_3 \s,
\ee
and $\th=\pi/2$ the solutions that we get are for complex $y$.
In the Appendix B we present several ansatze that do not have acceptable solutions. Some of these because are not consistent with the Sasaki-Einstein constraints and some other because  are complex.

We can use more complicated ansatze, for example by switching off the $\sigma$ dependence only on one $U(1)$ angle and keep it in the other two, but then the whole system of equations become complicated. However, it is interesting to consider the general ansatz
\ben\label{ansatz18}
\a=\o_1\t+m_1 \s,\qquad\phi = \o_2\t+m_2 \s,\qquad \psi =\o_3 \t + m_3\s,
\een
and set the constant angle equal to $\theta=\pi/2$. The system of equations is complicated and difficult to be solved, so the method we follow is to try to get rid off the most complicated equation \eq{eom2} by choosing appropriately the $\o_i,\,m_i$. We can consider
\ben
&&\o_1=s_3\,\o_3,\,\quad\o_2=s_2 \o_3,\,\quad\mbox{hence}\quad \,\o_1=\frac{s_3}{s_2}\,\o_2~,\\
&&m_1=t_3 m_3 ,\quad m_2=t_2 m_3,\quad \mbox{hence}\quad m_1=\frac{t_3}{t_2}\, m_2~.
\een
Then the equation \eq{eom2} is solved for
\be\label{lysh18set}
m_3^2=\o_3^2,\qquad  s_2^2= t_2^2,\qquad s_3= t_3~,
\ee
and we choose the set of solutions
\be\label{lysh18sol1}
s_3=t_3,\qquad s_2=-t_2,\qquad m_3=-\o_3 .
\ee
The remaining equations \eq{eom1} and \eq{vc1} are solved for
\ben\label{lysh18sol2}
a&=&\frac{t_2^4 (1+6 t_3)+16 t_3^2 (1+18 t_3)+8 t_3 (t_2+6 t_2 t_3)^2}{(1+6 t_3) \left(t_2^2+4 t_3 (1+6 t_3)\right)^2}~,\\
y&=&\frac{t_2^2+4 t_3}{t_2^2+4 t_3 (1+6 t_3)}~.
\een
This solution clearly satisfies \eq{yqy} as we see in the Figure 9. Moreover, there is no constraint on the winding numbers since they can take integer values. Hence we can calculate the momenta which in our case give
\be
J_\a=\sqrt{\l}\frac{4 \o_3 t_2^2 t_3}{t_2^2+4 t_3(1+6 t_3)},\qquad J_\a=-\frac{t_2}{t_3} J_\phi=-3 Q_R,\qquad J_\psi=0\, .
\ee
\begin{figure}\label{flysh18}
\includegraphics[width=55mm]{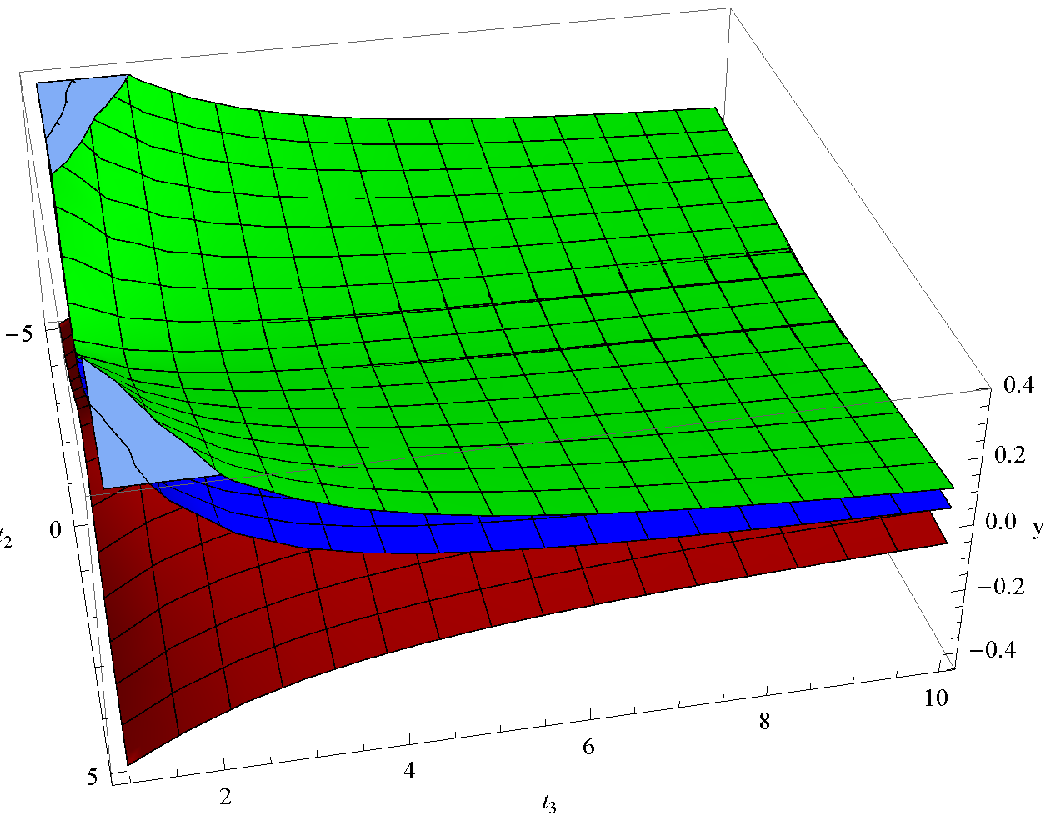}
\includegraphics[width=70mm]{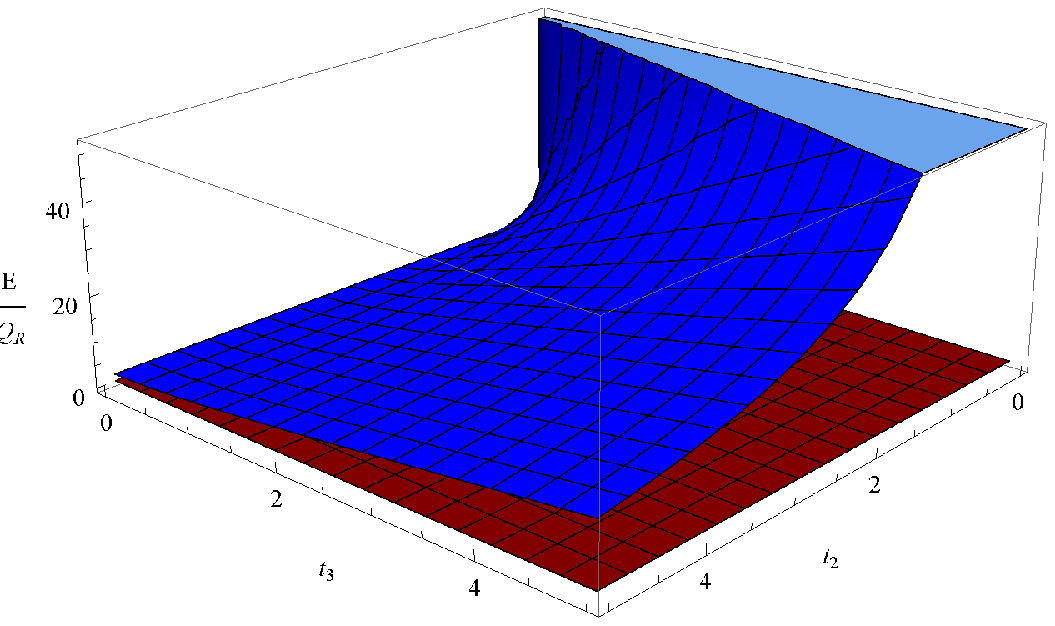}
\caption{{\small In the first plot we notice that our solution $y$ satisfies the inequality \eq{yqy} since it is always between $y_{q\pm}$ for all values of $t_2$ and $t_3$. We plot only a region of the whole space in order to be able to see the behavior of the functions clear. The behavior is similar to the whole space for $t_3<0$. In the second plot is presented the $E/Q_R$ with running variables $t_2$ and $t_3$. We see that the lower bound of the function is 3 and greater than $3/2$.}}
\end{figure}
The goal is to express the energy in terms of the momenta, using the second Virasoro equation which reads
\be
\k^2=\frac{16  t_2^2 t_3^2 \o_3^2}{t_2^2+4 t_3 (1+6 t_3)}~.
\ee
Hence the energy in terms of the R-charge and $t_{2,3}$ is given by
\be
E=3 \sqrt{\frac{t_2^2+4 t_3(1+6 t_3)}{t_2^2}} |Q_R|~.
\ee
The factor of proportionality is plotted in Figure 9 and as expected  is always greater than $3/2$, with its lower bound to be equal to $3$. Also one can see that as $t_3$ increases $a$ increases as well the energy of the solution. In this case however the solutions $y$ seem not to approach the $y_{q\pm}$ for generic $t_i$, and hence we did not expect to see the energy of this solution to approach the BPS one.

Here we have constrained our ansatz significantly. It would be interesting to check if there are any solutions that relate $s_i,\,t_i$ in the most minimal way or even better not at all. It is obvious however that this task is computationally very demanding, and if solvable possibly will lead to a complicated dispersion relation.



\section{Discussions}

In this paper, we analyze several string solution in $Y^{p,q}$ backgrounds. Initially we consider a string extending in the $\theta$ direction, which is not isometry of the manifold and simultaneously allow spinning along the $U(1)$ directions. Later we fix the angle $\theta$ and generalize the motion along the $U(1)$ directions where we find several point-like and extended string solutions.

In the energy-spin relations of some of our solutions we notice a common feature. If we can place the string close to the boundaries $y_{q\pm}$ for the maximum allowed value of the parameter $a$,  specified by the inequality \eq{yqy}, their energy  approaches the energy of the BPS solutions. This behavior of these specific string configurations is independent of the exact string motion in the other directions and can occur for different values of the parameter $a$. The only requirement is that for the maximum value of  $a$, the solution $y$ approaches $y_{q\pm}$, and this is always true when the solutions $y(a)$ constrained by the inequality \eq{yqy}.
We found several string solutions with the above properties. For example the energy \eq{lysh5energy} corresponds to a string that can be placed close to $y_{q\pm}$ for the maximum allowed value $a\rightarrow 1/2$\footnote{For the parameter $a$, defined by \eq{apq}, might not be possible to get the exact value of $1/2$, but there exist manifolds with parameter $a$ very close to this value. We expect that one can construct solutions such that for certain parameters $a$ which correspond exactly to specific $Y^{p,q}$ manifolds, the generic non-BPS energy-spin relation becomes equal to the BPS one.}. The string here is extended in $\a$ and $\psi$ directions and allowed to spin only along $\a$ direction. There are other cases, for example the energy \eq{energeialysh8} of the point-like string in the $\a$ and $\phi$ directions, or the energy \eq{energeia1} which corresponds to a string extended in $\theta$ and spinning along the $\a$ direction, where their upper bound of $a$, is for $a\rightarrow 1$\footnote{However notice that the limit $a\rightarrow 1$ is not so interesting as $a\rightarrow 1/2$. The reason is that in this limit the Sasaki-Einstein spaces approach the $S^5$. Nevertheless, it is definitely worthy to mention the behavior of the energy solutions even there.}. Hence for the string solutions we are referring here, the maximum value of $a$ specified such that the Sasaki-Einstein constraints are satisfied, and at the same time corresponds to placing the string close to $y_{q\pm}$. The energy of these solutions at this region approaches the energy of the BPS point-like string solutions. This also implies that the energy in these cases is a decreasing function of $a$.

Moreover, it is clear that the string solutions satisfy the Sasaki-Einstein constraints have energy above the BPS one. In other words, the constraints of the manifold ensure that the non-BPS string solutions can not have energy lower of the BPS solutions energy for any values of the parameters. This is expected and a characteristic example is the string \eq{lysh5}, where its solution satisfy the inequality \eq{yqy} for $a<1/2$ as we already mention before. For $a>1/2$ the energy of the solution becomes smaller than the BPS one, but due to Sasaki-Einstein constraints the solutions for these values of $a$ are not acceptable.

It would be very interesting to relate the energy of the solutions found here with the conformal dimension of the corresponding  operators in field theory in a way similar done in the original AdS/CFT conjecture. These operators should be constructed only with the use of matter bifundamental fields, since our strings do not have any AdS angular momenta.
Notice that when studying BPS quantities in these spaces, the explicit knowledge of the Calabi-Yau metric is not required and is enough to use only the general characteristics of the Calabi-Yau metric. However for the non-BPS cases we expect that the full metric is needed as our results suggests. Furthermore, of a special interest is how one can see that the corresponding field theory operators to our solutions, become BPS at certain limits in particular quivers. In \cite{benvenutiY} it has been found that there exist  points on the conformal manifold where some coefficients of the superpotential vanish and the chiral ring is enhanced. Hence, at these points some holomorphic operators becomes BPS and in string side this includes the extended semiclassical strings. For example in particular quivers, for some semiclassical strings moving on the round sphere, and localized at the south pole of the squashed sphere, meaning $y=y_{q-}$, one can find the corresponding generic operator. At the point where some terms of the superpotential vanish, the operator become BPS and satisfies the relation $\D=3/2 Q_R$. This is similar to what happens to our string solutions, and would be very interesting to investigate further the corresponding field theory picture.

On the other hand, we find string solutions that their energies do not approach the BPS energy, even in cases that the position of the string in $y$ direction approaches  $y_{q\pm}$ for the maximum value of the parameter $a$. An example of this kind is the solution for the string parametrized by the relations \eq{ansatz10}. In that case one can argue that the energy is not defined well at $a\rightarrow 1$. However, we can see that even for the string configuration \eq{ansatz8}, the energy does not approach the BPS one in the lower bound of the parameter $a$\footnote{The energy of this solution does it for the maximum value of $a$ as we mention above.} which corresponds to $y\rightarrow y_{q+}$. Hence the fact that we can place a string close to the $y_{q\pm}$ for a boundary value of the allowed interval of the parameter $a$ specified by the Sasaki-Einstein constraints, does not always mean that  its energy will approach the BPS one.

Notice also that the dispersion relations of the strings analyzed, depends always on the parameter $a$, hence the manifold considered. One could think the possibility to relate this dependence to the volume of the Sasaki-Einstein manifolds. However,  there is no relation between them, and this can be seen easily by noticing that our energy-spin relations have monotonic behavior with respect to $a$, while the geometry with the largest volume of the Sasaki-Einstein manifolds occur for $Y^{2,1}$ which corresponds to $a\simeq 0.387$. The only known string solution that its energy does not depend on the exact manifold considered is the point-like BPS, and this reflect to the fact that when calculating BPS quantities in Sasaki-Einstein spaces, the exact knowledge of the metric is not required.

Our method of finding semiclassical string solution in $Y^{p,q}$ manifolds can be directly applied to the  cohomogeneity two manifolds $L^{p,q,r}$. For these manifolds however, more parameters are involved and it is more complicated to check analytically whether or not the relevant solutions, satisfy the manifold constraints.

It is also very interesting to look at the effect of the $\beta$ deformations \cite{LM} on the string solutions on these manifolds. So far in the deformed toric quiver theories have been analyzed the BPS dual giant gravitons  \cite{zaffaronibse} and the spiky strings \cite{rashkovt11}. An interesting result found in the first paper where the determinant of the metric of the two torus created by the angles on which the TsT  \footnote{One can deform the original backgrounds with a global $U(1)\times U(1)$ symmetry to the $\beta$-deformed ones, by applying a T-duality on one $U(1)$ angle, a shift on a second U(1) angle where the deformation parameter enters and finally a T-duality on the initial angle \cite{LM,frolovtst}.} applied has to vanish in the $D3$ dual giant graviton solution and forces it to live on the edges of the corresponding polyhedron. It is interesting to see how and if the string BPS solution will be modified in the deformed Sasaki-Einstein background. Furthermore, it seems that for extended strings in these deformed spaces  some new features appear. It turns out that the deformation parameter enter in some solutions in a way that allows them to become real only in the deformed space and while switching off the parameter, the solutions become complex. The analysis of BPS and extended string solutions in deformed toric quiver theories  is a work in progress \cite{giataganasbse}.

Finally, our method of finding string solutions in Sasaki-Einstein manifolds could be used also to construct minimal surfaces on these manifolds that should correspond to Wilson loop operators. Although the Wilson loop operator in the toric theories is unknown and not so much progress is done, it is known that the UV divergences of the expectation value are canceled with the use of the Legendre transform \cite{giataganaswluv} and in other words the minimal surface conjecture for the Wilson loop operator in the original AdS/CFT  should remain intact and the Wilson loops
in Sasaki-Einstein dualities should be well defined. %

\startappendix
\Appendix{$Y^{p,q}$ Background}
The Sasaki--Einstein metrics $Y^{p,q}$ \eq{se1} on $S^2\times S^3$ can take the local compact form
\be\label{metricse2}
 d s^2 =d s^2(B)+w(y)[d \alpha+A]^2~,
\ee
where the functions $w(y),\,q(y)$ and $f(y)$ are given in section 2.
To make the space $B$ a smooth complete compact manifold we should fix the coordinates appropriately \cite{gauntletts2s3}. The parameter $a$ is restricted by the inequality \eq{ai}.
Moreover to make  the base
$B_4$ an axially squashed $S^2$ bundle
over the round $S^2$, the ranges of the
coordinates $(\theta,\phi,y,\psi)$ should be chosen as in section 2.

The quasi-regular Sasaki-Einstein manifolds which have $y_{q\pm}$ rational, have the property that the volume of these manifolds having a rational relation to the volume of the $S^5$. However,  the rationality of $\xi$ , defined in \eq{xi}, can be achieved even in cases that the two roots are irrational. In the following we collect some very useful equations of the Sasaki-Einstein constraints where it can be also seen how the irregular metrics arise.

The three roots of cubic satisfy
\ben\label{yprop}
y_{q+}+y_{q-}+y_3=3/2,\quad
y_{q+}y_{q-}+y_{q+}y_3+y_{q-}y_3=0,\quad
2 y_{q+}y_{q-}y_3=-a
\een
and also can be expressed in terms of $p,\,q$
\ben
y_{q\pm}=\frac{1}{4p}(2 p \pm 3 q-\sqrt{4 p^2-3 q^2}),\quad
y_3=\frac{1}{4p}(2 p +2\sqrt{4 p^2-3 q^2})~.
\een
Using the above expressions the negative root $y_{q-}$ can be written in terms of $\xi$
\ben
y_{q-}=\frac{1}{2}(1-\xi-\sqrt{1-\frac{\xi^2}{3}})~.
\een
Since $y_{q-}$ is the root of the cubic,  $a$ can be expressed in terms of $\xi$
\ben\label{axi}
a=\frac{1}{18} \left(9-3 \sqrt{9-3 \xi^2}+4 \xi^2 \sqrt{9-3 \xi^2}\right)~,
\een
and in order to ensure that $y_{q+}$ is the smallest positive root we constrain $\xi$ to the range $0<\xi<3/2$.

Moreover, we can express $a$ in terms of $p,\,q$ using \eq{xi}
\be\label{apq}
a=\frac{1}{2}-\frac{p^2-3q^2}{4 p^3}\sqrt{4 p^2 -3 q^2}~,
\ee
then the period of $\a$ is given by $2 \pi l$ where
\ben\label{aper}
l=\frac{q}{3q^2-2 p^2+p(4p^2-3 q^2)^{1/2}}~,
\een
or in a more compact form
\be
l=-\frac{q}{4 p^2 y_{q+} y_{q-}}~ ,
\ee
which is always positive since $y_{q-}$ is negative. Finally the volume $Y^{p,q}$ is given by
\ben
Vol(Y^{p,q})=\frac{q(2 p +\sqrt{4 p^2-3 q^2})l \pi^3}{3 p^2}
\een
and is bounded by
\be
Vol(T^{1,1}/\mbox{\cZ}_p)>Vol(Y^{p,q})>Vol(S^5/\mbox{\cZ}_2 \times \mbox{\cZ}_p).
\ee

\Appendix{Other strings configurations with three angles activated}

There are several string ansatze for string's motion with activated all three $U(1)$ angles and $\th=\pi/4$, that do not solve the equations of motions and the Virasoro constraints. Even in the case that these are solved, the solutions are not consistent with the Sasaki-Einstein constraints and the periodicity boundary conditions. One main reason that creates problems with the integer values of the winding numbers, is that although the choice of the angle simplifies the equations, since the $\cos\th$ and $\sin\th$ are equal, it inserts in the equations the irrational $\sqrt{2}$, and as a result this enters often between the relation of the winding numbers. Of course one can simplify the relation a lot by setting $\th=\pi/2$ as we already done to solve some very complicated equations. But in general we avoid to do that since this value will make equal to zero many terms of the equations of motion and the Virasoro constraints and hence our solutions do not capture the full dynamics of the theory.

We point out that the motion of the string in the Sasaki-Einstein manifolds for strings spinning according to the following relations\footnote{Where the angle $\th$ is fixed to $\pi/4$.} is impossible:
\ben\label{anaf335151}\nn
&\a=m_1 \s,&\quad\phi = \o_2 \t+ m_2 \s,\quad \psi = m_3\s,\\\nn
&\a=m_1 \s,&\quad\phi = m_2 \s,\qquad\quad\,\,\,\,  \psi =\o_3 \t + m_3\s,\\\nn
&\a=m_1 \s,&\quad\phi = m_2 \s,\qquad\quad\,\,\,\, \psi=\o_3 \t,\\\nn
&\a=m_1 \s,&\quad\phi = \o_2 \t,\qquad\quad\,\,\,\,\, \psi = m_3\s,\\\nn
&\a=m_1 \s,&\quad\phi = \o_2\t,\qquad\quad\,\,\,\,\, \psi = \o_3\t\\\nn
&\a=\o_1 \t,&\quad\phi = m_2\s,\qquad\quad\,\,\, \psi = m_3 \s,\\\nn
&\a=\o_1 \t,&\quad\phi = \o_2\t,\qquad\quad\,\,\,\,\, \psi = m_3 \s,\\\nn
&\a=\o_1 \t,&\quad\phi = m_2\s,\qquad\quad\,\,\,\, \psi = \o_3 \t,
\een
since the equations \eq{eom1}, \eq{eom2}, \eq{vc1} do not give new solutions which satisfy the Sasaki-Einstein constraints and the periodic boundary conditions.

But for example by setting the $\theta$ angle equal to $\pi/2$ and considering the ansatz
\be
\a=m_1 \s,\qquad \phi = m_2 \s,\qquad \psi =\o_3 \t~,
\ee
we get the simple solution
\be
\o_3=\frac{\sqrt{3}}{2} m_2,\qquad
y=1-\sqrt{1-a}~.
\ee
Which has only a non zero angular momentum
\be
J_\psi=\frac{Q_R}{2}=\frac{\sqrt{1-a}~ m_2}{3 \sqrt{3}}
\ee
and the second Virasoro constraint has the simple form
\be
\k^2=4 \left(1-\sqrt{1-a}\right) m_1^2+\frac{1}{4} \sqrt{1-a} m_2~.
\ee
Using the above relation we get the final energy in terms of the momenta
\be
E=\sqrt{\frac{27 Q_R^2}{4 \sqrt{1-a}}+4 \left(1-\sqrt{1-a}\right) m_1^2}\, .
\ee
We again notice here that the factor in front of $Q_R$ is greater than $3/2$.

This was an example of how the choice of the angle $\theta$ can simplify the analysis, but at the same time we might be losing significant part of dynamics of the background.

\section*{Acknowledgements}

I would like to thank Chong-Sun Chu, Robert de Mello Koch and Antal Jevicki for useful discussions.
Part of this work was done while the author was at Centre of Particle Theory, Durham University, UK.
The research of the author is supported by a SARChI postdoctoral fellowship.

\end{document}